%% file: supercurrent.tex
\documentclass[12pt, a4paper]{article}
\pdfoutput=1

\usepackage[utf8]{inputenc}
\usepackage{microtype}
\usepackage[a4paper]{geometry}
\usepackage{graphicx, adjustbox}
\usepackage[mcite,backend=biber, maxbibnames = 99, style=numeric-comp, doi=false, isbn=false, url=false, eprint=false, sorting=none]{biblatex}
\usepackage{amsmath, amssymb, amsthm, mathrsfs}
\usepackage{multirow, booktabs, caption, rotating}
\usepackage{tikz}
\usetikzlibrary{shapes,arrows}

\renewbibmacro{in:}{%
  \ifentrytype{article}{}{\printtext{\bibstring{in}\intitlepunct}}}

\DeclareFieldFormat[article]{title}{}
\DeclareFieldFormat[incollection]{title}{}
\numberwithin{equation}{section}

\begin{document}

\begin{center}
\vspace*{2.5cm}
{\Large\bf Chiral Four-Dimensional Heterotic Covariant Lattices}\\
\vspace*{1.5cm}
{\large  Florian Beye$^{1}$\footnote{Electronic address: fbeye@eken.phys.nagoya-u.ac.jp} }\\
\vspace*{1.0cm}
{\it $^1$Department of Physics, Nagoya University,\\Fur\={o}-ch\={o}, Chikusa-ku, Nagoya 464-8602, Japan}\\
\vspace*{1.5cm}

\begin{abstract}
In the covariant lattice formalism, chiral four-dimensional heterotic string vacua are obtained from certain even self-dual lattices which completely decompose into a left-mover and a right-mover lattice. The main purpose of this work is to classify all right-mover lattices that can appear in such a chiral model, and to study the corresponding left-mover lattices using the theory of lattice genera. In particular, the Smith-Minkowski-Siegel mass formula is employed to calculate a lower bound on the number of left-mover lattices. Also, the known relationship between asymmetric orbifolds and covariant lattices is considered in the context of our classification.
\end{abstract}

\end{center}
\vspace*{-1.0cm}
\newpage
\tableofcontents
\vspace*{0.5cm}


\section{Introduction}
String theory may eventually provide a consistent quantum-mechanical unification of elementary particle physics with gravity. This prospect has spawned countless efforts on string model building and already led to promising results. For example, orbifold compactifications of the heterotic string~\mcite{Orbifolds,*Dixon:1985jw,*Dixon:1986jc} successfully reproduce many properties of the standard model of particle physics, such as its gauge symmetry group and three chiral generations of matter~\mcite{HetOrbModels,*Lebedev:2006kn,*Lebedev:2008un}. Recently, also asymmetric orbifolds~\mcite{AsymOrbifolds, *Narain:1986qm, *Narain:1990mw} were considered for model building~\cite{Beye:2013moa, Beye:2013ola}. 

The covariant lattice formalism \mcite{CovLat, *Lerche:1986he,*Lerche:1986cx} provides another exact CFT construction of four-dimensional heterotic string vacua (\cite{Lerche:1988np} for a review). In this formalism, all internal world-sheet degrees of freedom are expressed in terms of free bosons with periodic boundary conditions. These boundary conditions are encoded in a lattice $\Gamma_{22,14}$ -- the covariant lattice, which must be even and self-dual due to modular invariance and obeys further constraints from world-sheet supersymmetry. Additionally requiring a chiral four-dimensional spectrum renders the number of possible vacua finite. Some covariant lattice models can also be obtained from other constructions. For example, it is known that by bosonization certain asymmetric $Z_N$ orbifolds are equivalent to a covariant lattice theory~\mcite{AsymOrbCL, *Schellekens:1987ij, *Schellekens:1988ag}. Also, there are some overlaps with free fermionic constructions and Gepner models, as indicated in  \cite{Lerche:1988np}. As of today, only few models were constructed explicitly using the covariant lattice formalism (see e.g.~\cite{Nilsson:1989pn}), and a complete classification has only been achieved in ten spacetime dimensions~\cite{Lerche:1986ae}. In eight dimensions, 444 chiral models were found to exist~\cite{Balog:1989xd}. The phenomenologically interesting case of four dimensions was considered in~\cite{Balog:1987vf}, however their treatment is too restrictive and does not cover all chiral models.

A special feature of \textit{chiral} four-dimensional covariant lattice models is that $\Gamma_{22,14}$ contains a sublattice of the form ${(\Gamma_{22})}_\text{L} \oplus \overline{(\Gamma_{14})}_\text{R}$. The purpose of this paper is to provide a (computer aided) classification of right-mover lattices $(\Gamma_{14})_\text{R}$ which solve the constraints imposed by world-sheet superconformal invariance and chiralness. The result is that there are in total 99 such lattices, and 19 of them lead to $\mathcal{N} = 1$ spacetime supersymmetry. In order to construct a complete model, one has to combine one of these lattices with an appropriate left-mover lattice $(\Gamma_{22})_\text{L}$. In fact, due to modular invariance, all lattices $(\Gamma_{22})_\text{L}$ that can be combined with a chosen $(\Gamma_{14})_\text{R}$ constitute a genus, so the very well-developed theory of lattice genera can be used to study them (an introduction on the subject can be found e.g. in~\cite{SPLAG}). Most importantly, a genus $\mathcal{G}$ contains only finitely many lattices, and a lower bound on their number $\vert \mathcal{G} \vert$ can be calculated by means of the Smith-Minkowski-Siegel mass formula (refer to \cite{Conway:1988} for details). Furthermore, there are computational methods which allow the explicit construction of all lattices in a genus. However, these are only practicable when $\vert \mathcal{G} \vert$ is reasonably small. In this work, some of the relevant genera $\mathcal{G}_\text{L}$ of left-mover lattices are enumerated explicitly (one in particular corresponds to a certain class of $Z_3$ asymmetric orbifolds) and for some other genera, a lower bound on $\vert \mathcal{G}_\text{L} \vert$ is calculated. These lower bounds suggest that there are at least $O(10^{10})$ four-dimensional covariant lattice models realizing $\mathcal{N}=1$ spacetime supersymmetry. Finally, this work indicates that all chiral covariant lattice models are related to certain Narain-compactified $\mathcal{N} = 4$ theories by shift-orbifolding. It is further shown that some models cannot be obtained from twist-orbifold constructions similar to those in~\cite{AsymOrbCL}.

The paper is organized as follows. In Section 2, we first review some relevant aspects of the covariant lattice construction and then formulate the constraints implied from requiring a chiral spectrum. Section 3 treats the classification of chiral covariant lattices and discusses the relationship with certain orbifold constructions. Section 4 is devoted to conclusions and shows possible implications of the results.


\section{The Covariant Lattice Formalism}
In this section we briefly introduce the covariant lattice (or bosonic supercurrent) formalism. We also derive the constraints from world-sheet superconformal invariance and from the requirement of a chiral four-dimensional spectrum. For a more detailed introduction, please refer to the review \cite{Lerche:1988np}.


\subsection{Bosonic Realizations of Supersymmetry}\label{ss:covlat}
As a consequence of anomaly cancellation, any four-dimensional heterotic string theory requires an internal unitary CFT with central charges $(c_\text{L}, c_\text{R})=(22,9)$. Here, we follow the approach of the bosonic supercurrent formalism \cite{CovLat} and consider only internal CFTs that are realized entirely in terms of free chiral bosons with periodic boundary conditions. These boundary conditions are encoded in an even Lorentzian lattice $\Gamma_{22,9}^{\text{int}}$ of signature $(22,9)$. Moreover, it is convention to bosonize the four Neveu-Schwarz-Ramond fermions $\psi^\mu$ as well as the $\beta\gamma$-ghosts, and map them to a right-mover ``spacetime'' $SO(10)$ root lattice $D_5^\text{st}$ via the bosonic string map~\mcite{BosStringMap, *Casher:1985ra, *Englert:1986na}. Then, modular invariance is guaranteed if $D_5^\text{st}$ and $\Gamma_{22,9}^{\text{int}}$ can be glued together to an even self-dual lattice $\Gamma_{22,14}$, i.e.:
\begin{align}\label{cdecomp}
	\Gamma_{22,14} \supset \Gamma_{22,9}^{\text{int}} \oplus \overline{D_5^\text{st}} \,.
\end{align}
The lattice $\Gamma_{22,14}$ is called covariant lattice and is of signature $(22,14)$. Here and in the rest of the paper, $\overline{\Lambda}$ denotes the lattice which is identical to $\Lambda$ except that its inner product is amended by an additional minus sign. 

The covariant lattice obeys additional constraints due to supersymmetry: since the internal right-mover $c_\text{R} = 9$ CFT is supersymmetric, there has to exist a supercurrent $G(z)$ that obeys the $N=1$ super-Virasoro algebra,
\begin{align}
	T(z) T(w) & \sim \frac{c_\text{R}/2}{(z-w)^4} + \frac{2 T(w)}{(z - w)^2} + \frac{\partial T(w)}{(z - w)}\\ 
	\label{cweightG} T(z) G(w) & \sim \frac{\tfrac{3}{2} G(w)}{(z - w)^2} + \frac{\partial G(w)}{(z - w)}\\ 
	\label{cGG} G(z) G(w) & \sim \frac{2c_\text{R}/3}{(z-w)^3} + \frac{2 T(w)}{(z - w)} \,.
\end{align}
In our case, the internal right-mover CFT is that of nine free chiral bosons $X^i(z)$, for which the the energy-momentum tensor has the standard form
\begin{align}
	T(z) & = -\frac{1}{2} :\partial X(z) \cdot \partial X(z): \, .
\end{align}
Obviously, we also need to express the supercurrent in terms of the $X^i(z)$. The OPE~(\ref{cweightG}) states that $G(z)$ is a primary field of conformal weight $3/2$. This condition is solved by the following expression:
\begin{align}\label{scdef}
	G(z) & =\;:\sum_{s^2 = 3} A(s) e^{is\cdot X(z)} \varepsilon(s, \hat p) + \sum_{r^2 = 1} i B(r) \cdot \partial X(z) e^{ir \cdot X(z)} \varepsilon(r, \hat p):\,.
\end{align}
Here, one sums over two sets of vectors $r$ and $s$ of norm $1$ and $3$, respectively. Also, the coefficients $B^i(r)$ are subject to a transversality condition,
\begin{align}
r \cdot B(r) = 0\,.
\end{align}
Moreover, in equation (\ref{scdef}), 
\begin{align}
	\hat p^i = \frac{1}{2\pi} \oint dz \, \partial X^i (z) 
\end{align}
denotes the ``center-of-mass" momentum operator and $\varepsilon(s,t)$ is a cocycle that has to be introduced to ensure correct statistics for $G(z)$. The constraints from the OPE~(\ref{cGG}) on the coefficients $A(s)$ and $B^i(r)$ shall not be considered in full generality here. Instead, we will discuss a special case in Subsection~\ref{ss:chiral}, where we additionally require a chiral four-dimensional spectrum.

\subsubsection*{Constraint Vectors}
Due to an additional consistency condition, for each vector $r$ and $s$ that appears in (\ref{scdef}) with non-zero coefficient, the covariant lattice must contain all vectors of the form
\begin{align}
(0;r,v)\text{ and }(0;s,v)\,,
\end{align}
called constraint vectors of the supercurrent~\cite{Lerche:1988np}. In above notation, the first entry belongs to the 22-dimensional left-mover subspace and the second one to the internal nine-dimensional right-mover subspace. The third entry corresponds to the subspace spanned by $\overline{D_5^\text{st}}$ and $v$ denotes a weight in the vector conjugacy class of $D_5$. 

Since the covariant lattice is an even lattice and $v^2$ is an odd integer, an immediate consequence is that the vectors $r$ and $s$ span an odd integral lattice. In the following, we denote this ``supercurrent lattice" by $\Xi$, and sums in expressions such as~(\ref{scdef}) are to be interpreted as sums over vectors in $\Xi$. 

\subsubsection*{The Cocycle}
We now need to discuss some properties of the 2-cocycle, $\varepsilon: \Xi \times \Xi \mapsto \mathbb{C}^\times$ (here, $\mathbb{C}^\times$ shall denote the non-zero complex numbers). It is subject to the conditions
\begin{align}
\label{coc1} \varepsilon(s, t)\,\varepsilon(s + t, u) & = \varepsilon(s, t + u)\,\varepsilon(t, u) \\
\label{coc2} \varepsilon(s, t) & = (-1)^{s\cdot t + s^2t^2} \, \varepsilon(t, s) \,,
\end{align}
for all $s,t,u \in \Xi$. These conditions are preserved if the cocycle is multiplied by a coboundary $\delta\eta(s,t)$, i.e. 
\begin{align}\label{cocG}
\varepsilon(s,t)^\prime = \varepsilon(s,t) \,\eta(s)\,\eta(t)\,\eta(s+t)^{-1}
\end{align}
is also a valid cocycle. Then, by choosing $\eta$ appropriately, the cocycle can be gauged to satisfy
\begin{align}
\label{coc3} \varepsilon(0, s) & = 1\\
\varepsilon(s, -s) & = 1\\
\varepsilon(s, t) &= \varepsilon(t, -s - t) = \varepsilon(-s-t, s)\\ 
\label{coc6} \varepsilon(s, t)^\ast & = \varepsilon(-t, -s) = \varepsilon(s, t)^{-1}\,,
\end{align}
for all $s,t \in \Xi$ (cf. also \cite{Goddard:1986bp}). These properties are assumed in the rest of the paper. 

A particularly useful way to obtain an explicit realization is as follows. First, define $\varepsilon(b_i, b_j)$ for a basis $\lbrace b_1, \ldots, b_9\rbrace$ of $\Xi$ in such a way that the conditions
\begin{align}
\varepsilon(b_i, b_i) & = 1\\
\varepsilon(b_i, b_j)\, \varepsilon(b_j, b_i) & = 1\\
\varepsilon(b_i, b_j) & = (-1)^{b_i\cdot b_j + b_i^2b_j^2}  \, \varepsilon(b_j, b_i) 
\end{align}
are satisfied, and then consider the linear continuation to $\Xi \times \Xi$. One can show that this implies the other properties.

\subsubsection*{Spacetime Supersymmetry}
Finally, we want to give a note on spacetime supersymmetry, which is particularly straightforward to read off in covariant lattice models. Spacetime supersymmetry is attained if the covariant lattice contains vectors that extend $D_5^\text{st}$ to one of the exceptional root lattices $E_6^\text{st}$, $E_7^\text{st}$, and $E_8^\text{st}$. These correspond to the cases $\mathcal{N} = 1$, $\mathcal{N} = 2$ and $\mathcal{N} = 4$, respectively.

The $\mathcal{N}=4$ case deserves some extra mentioning. Then, $\Gamma_{22,14}$ contains an $\overline{E_8^\text{st}}$-sublattice, which is self-dual by itself. Hence, the covariant lattice can be written as an orthogonal sum,
\begin{align}
	\Gamma_{22,14} = \Gamma_{22,6} \oplus \overline{E_8^\text{st}}\,,
\end{align}
where $\Gamma_{22,6}$ is self-dual and known as Narain lattice. This scenario describes the most general toroidal compactification including background fields~\mcite{NarainLat, *Narain:1985jj, *Narain:1986am}. More details on spacetime supersymmetry in covariant lattice theories can be found in the review~\cite{Lerche:1988np}.


\subsection{Covariant Lattices for Chiral Models}\label{ss:chiral}
Now, we require that the four-dimensional effective theory has a chiral spectrum. Then, it can be shown that chiralness is spoiled if the supercurrent lattice $\Xi$ contains vectors $r$ of norm 1~\cite{Lerche:1988np}. In that case, $D_5^\text{st}$ is enhanced to some $D_{5+k}^\text{st} \supset D_5^\text{st}$, which causes massless fermion matter to appear in vector-like pairs. Hence, in the following we assume the absence of norm 1 vectors and use the ansatz
\begin{align}\label{ansatz}
G(z) & =\;:\sum_{s^2 = 3} A(s) e^{is\cdot X(z)} \varepsilon(s, \hat p): \,.
\end{align}
Note that the absence of norm 1 vectors is only a necessary condition for chiralness and it is nonetheless possible to obtain non-chiral models this way.

Until now, we did not consider equation (\ref{cGG}). So, the next step is to calculate the $GG$-OPE using the ansatz in equation~(\ref{ansatz}) and compare it with the r.h.s. of equation~(\ref{cGG}). Then, one obtains the following system of quadratic equations in the coefficients $A(s)$:
\begin{align}
\label{maineq1} \sum_{s^2 = 3} \vert A(s) \vert^2 s^i s^j & = 2 \delta^{ij} \\
\label{maineq2}\sum_{\substack{s^2 = t^2 = 3 \\ s + t = u}} A(s)A(t) \varepsilon(s,t) & = 0 \;\;\text{for all } u^2 = 4 \\
\label{maineq3}\sum_{\substack{s^2 = t^2 = 3 \\ s + t = u}} A(s)A(t) \varepsilon(s,t)\, (s^i - t^i) & = 0 \;\;\text{for all } u^2 = 2 \,.
\end{align}
Here, we also imposed a hermiticity condition,
\begin{align}
	G(z)^\dagger = (z^\ast)^{-3}G(1/z^\ast)\,,
\end{align}
implying $A(s)^\ast = A(-s)$. A generalized version of equations~(\ref{maineq1})--(\ref{maineq3}) which includes also non-vanishing $B(r)$ can be found in~\cite{Balog:1989xd}.

\subsubsection*{Left-Right Decomposition of the Covariant Lattice}
As an immediate consequence of equation (\ref{maineq1}), the supercurrent lattice $\Xi$ must completely span the nine-dimensional space it resides in. Then, the constraint vectors $(0; s, v)$ generate a negative-definite 14-dimensional sublattice $\overline{\Gamma}_{14}$ of $\Gamma_{22,14}$. We also define ${(\Gamma_{22})}_\text{L}$ as the 22-dimensional lattice of vectors belonging to the orthogonal complement of $\overline{\Gamma}_{14}$ in $\Gamma_{22,14}$, and similarly $\overline{(\Gamma_{14})}_\text{R}$ as the lattice of vectors in the orthogonal complement of ${(\Gamma_{22})}_\text{L}$ in $\Gamma_{22,14}$. Clearly, $\Gamma_{14} \subseteq {(\Gamma_{14})}_\text{R}$, and we obtain the following decomposition:
\begin{align}\label{chiraldecomp}
	\Gamma_{22,14} \supset (\Gamma_{22})_\text{L} \oplus \overline{(\Gamma_{14})}_\text{R} \,.
\end{align}
This decomposition does not contain self-glue, i.e. all elements of $\Gamma_{22,14}$ that belong to the $\mathbb{R}$-span of $\overline{(\Gamma_{14})}_\text{R}$ also lie in $\overline{(\Gamma_{14})}_\text{R}$, and those belonging to the $\mathbb{R}$-span of ${(\Gamma_{22})}_\text{L}$ also lie in ${(\Gamma_{22})}_\text{L}$. In the following, we always assume that decompositions such as in (\ref{chiraldecomp}) are self-glue free, and, that the lattice dimensions on the l.h.s. and on the r.h.s. of the inclusion relation match.

\subsubsection*{Symmetries of the Supercurrent Equations}
Equations~(\ref{maineq1})--(\ref{maineq3}) possess several symmetries (also cf.~\cite{Lerche:1988np}). First, note that the internal $c_\text{R} = 9$ right-mover CFT always contains the Kac-Moody algebra spanned by the currents $\partial X^i(z)$. To these currents corresponds the $U(1)^9$ symmetry group which is infinitesimally generated by the $\hat p^i$. Then, if $G(z)$ as in (\ref{ansatz}) satisfies the super-Virasoro algebra, so does the conjugate
\begin{align}
	U(\xi) G(z) U(\xi)^\dagger = \;:\sum_{s^2 = 3} A(s) e^{is\cdot\xi} e^{is\cdot X(z)} \varepsilon(s, \hat p): \,,
\end{align}
where $U(\xi) = e^{i\xi\cdot\hat p}$ denotes an element of the symmetry group. One also verifies that if $A(s)$ solves~(\ref{maineq1})--(\ref{maineq3}), then so does $A(s) e^{is\cdot\xi}$. 

Whenever there exist norm 2 vectors in $\Xi$, the $U(1)^9$ symmetry is enlarged to a non-Abelian group by additional Frenkel-Kac currents. Then, the norm 3 vectors correspond to weights of a (in general reducible) representation of this non-Abelian group, and group transformations can be used to set some of the $A(s)$ to zero. In this case it may happen that the $A(s)$ become non-zero only on a sublattice $\Xi^\prime \subset \Xi$, which can then be considered more fundamental than $\Xi$. Besides these continuous symmetries, there may also be additional discrete symmetries (e.g. those induced by lattice automorphisms) that transform one solution into another. 

In any case, above symmetries only correspond to a change of basis, and as such produce physically equivalent supercurrents. However, one can not rule out the possibility that equations~(\ref{maineq1})--(\ref{maineq3}) allow for distinct solutions not related by a physical symmetry. Then, one obtains inherently different string vacua. These vacua share the same spectrum, but string amplitudes which necessitate picture changing may differ because the internal supercurrent enters the picture changing operator.


\section{Classification of Chiral Covariant Lattice Models}
As explained in the last section, a covariant lattice $\Gamma_{22,14}$ corresponding to a chiral four-dimensional model decomposes as in equation~(\ref{chiraldecomp}). This allows us to attack the classification of these models as follows:
\begin{enumerate}
\item Enumerate all possible ${(\Gamma_{14})}_\text{R}$ that are consistent with the constraints from world-sheet supersymmetry and permit a chiral four-dimensional spectrum.
\item For each ${(\Gamma_{14})}_\text{R}$ found, determine the set of lattices ${(\Gamma_{22})}_\text{L}$ that can appear alongside ${(\Gamma_{14})}_\text{R}$ in the decomposition (\ref{chiraldecomp}) while respecting modular invariance.
\item Consider all possibly inequivalent embeddings of the obtained ${(\Gamma_{22})}_\text{L} \oplus \overline{(\Gamma_{14})}_\text{R}$ in an even and self-dual lattice.
\end{enumerate}
In the following, we carry out the first of the above steps. The second step will be discussed in Subsection \ref{ss:LM}. The third step will not be considered in detail in this work.


\subsection{Classification of Right-Mover Lattices}\label{ss:RM}
In Subsection~\ref{ss:chiral} it was shown that a right-mover lattice ${(\Gamma_{14})}_\text{R}$ of a chiral covariant lattice contains a sublattice $\Gamma_{14}$ that is constructed from a nine-dimensional lattice $\Xi$ using the constraint vectors. Further, we argued that $\Xi$ has the following properties:
\begin{enumerate}
\item \textit{Basic}. $\Xi$ is positive definite, integral, generated by its norm $3$ vectors, and contains no norm $1$ vectors. 
\item \textit{Supersymmetry}. There exists a solution $A(s)$ to equations~(\ref{maineq1})--(\ref{maineq3}) on $\Xi$.
\end{enumerate}
We call a positive-definite lattice $\Xi$ that obeys above two properties \textit{admissible}. It is our aim to classify all admissible lattices of dimension nine. Such a classifications in turn serves as a classification of all possible lattices $\Gamma_{14}$. Then, any right-mover lattice ${(\Gamma_{14})}_\text{R}$ of a chiral covariant lattice $\Gamma_{22,14}$ must contain one of these $\Gamma_{14}$ as a sublattice.

An admissible lattice can further be reduced to certain fundamental building blocks. First, note that an orthogonal sum of admissible lattices is again admissible. One also verifies that the converse is true: if an orthogonal sum of several factors is admissible, then so is each factor. Moreover, we do not need to care about an admissible lattice which contains an admissible sublattice of the same dimension. These facts motivate the definition of the following properties:
\begin{enumerate}
\item \textit{Primitivity}. $\Xi$ is not isomorphic to an orthogonal sum $\Xi_1 \oplus \ldots \oplus \Xi_k, \,k>1$. 
\item \textit{Elementarity}. $\Xi$ is admissible and does not contain a strictly smaller admissible sublattice $\Xi^\prime \subset \Xi$ of the same dimension.
\end{enumerate}
Then, any admissible lattice $\Xi$ can be built from primitive elementary building blocks by orthogonal composition and gluing. In the following we classify these building blocks.

\subsubsection*{The Primitive Basic Lattices}
The first step is to classify all lattices satisfying above basic and primitivity properties. This is done by induction over $n = \dim(\Xi)$ up to $n = 9$.

At this point we need to discuss a rather important subtlety. First, recall that any lattice $\Lambda$ has a basis $\lbrace b_1, \ldots, b_n \rbrace$ with basis vectors $b_i \in \Lambda$ such that each $x \in \Lambda$ is represented by a unique integer linear combination of the $b_i$. Then, $n$ denotes the dimension of $\Lambda$. When we say that a lattice $\Lambda$ is spanned (or generated) by a finite set $\lbrace a_1, \ldots, a_N \rbrace$ of vectors $a_i \in \Lambda$, we mean that any $x \in \Lambda$ can be written as a, not necessarily unique, integer linear combination of the $a_i$. Now, one might hope that, for any such generating set, it is possible to choose a basis where all basis vectors $b_i$ belong to that generating set. While this is certainly true for vectors spaces, it is not so for lattices (consider, for example, the lattice $\mathbb{Z}$ and the generating set $\lbrace 2, 3\rbrace$). A counterexample that comes close to our situation has been found in~\cite{Conway:1995}. The lattice constructed there is generated by its vectors of minimal length, but does not possess a basis solely out of these minimal vectors. After this discovery such lattices have received some interest~\cite{Martinet:2012}: they were shown not to exist for $n \leq 9$, while for $n = 10$ an example was found.

Although here we are only interested in the case $n \leq 9$, our lattices are allowed to contain vectors of norm 2, so the theorem of~\cite{Martinet:2012} is not applicable. Thus, among the lattices that satisfy our basic properties there might exist ``pathological" lattices which do not possess a basis of norm 3 vectors. However, suppose $\Xi^\prime$ be such a pathological case. Then, there must exist a sublattice $\Xi \subset \Xi^\prime$ that is not pathological. Hence, we can at the moment restrict our classification to lattices which do have a basis of norm 3 vectors and care about the pathological cases afterwards. Now, the idea is to enumerate these lattices by constructing all possible Gram matrices, in a way similar to the ``lamination" process introduced in \mcite{MinVec, *Plesken:1985, *Plesken:1993}. For a lattice with basis $\lbrace b_1, \ldots, b_n \rbrace$, the Gram matrix is defined by
\begin{align}
G_{ij} = b_i \cdot b_j\,.
\end{align} 
It determines the lattice up to $O(n)$ rotations in the ambient space, but is not a basis independent quantity. In our case we assume the existence of a basis of norm 3 vectors, so we only need to consider Gram matrices where the diagonal elements are $G_{ii} = 3$. Now, suppose $G^\prime$ be such a gram matrix of a $(n+1)$-dimensional lattice which satisfies our basic properties. The lattice corresponding to the restricted Gram matrix $G = (G_{ij}^\prime)_{i,j \leq n}$ then must also fulfill these properties. Now, we can write $G^\prime$ as 
\begin{align}\label{matdecomp}
	G^\prime = \begin{pmatrix}
		G  & v \\
		v^T & 3
	\end{pmatrix} \,,
\end{align}  
where $v$ is a column vector. From positive definiteness it follows that $\det(G^\prime) > 0$. This translates into the following condition:
\begin{align}\label{posdefcond}
	v^T G^{-1} v < 3\,.
\end{align}
Since the $v_i$ are integers and $G$ is positive definite, there are only finitely many possible $v$ satisfying equation~(\ref{posdefcond}). 

The naive algorithm then goes as follows. Suppose we have a set $\mathscr{B}_n$ that contains a Gram matrix for each non-pathological lattice of dimension $n$ with the basic properties. Set $\mathscr{B}_{n+1} = \lbrace\rbrace$. Then, for each pair $(G, v)$ with $G \in \mathscr{B}_n$ and $v$ satisfying equation~(\ref{posdefcond}) do the following:
\begin{enumerate}
\item Construct the matrix $G^\prime$ as in~(\ref{matdecomp}). It is necessarily positive definite. 
\item If the lattice corresponding to $G^\prime$ contains norm 1 vectors, continue to the next pair $(G, v)$.
\item Otherwise, check whether $\mathscr{B}_{n+1}$ already contains a Gram matrix that is equivalent to $G^\prime$ by a change of basis. If not, replace $\mathscr{B}_{n+1}$ by $\mathscr{B}_{n+1} \cup \lbrace G^\prime \rbrace$.
\end{enumerate}
After completion, $\mathscr{B}_{n+1}$ contains a Gram matrix for each non-pathological lattice of dimension $n+1$ satisfying our basic properties. The algorithm is initialized with $\mathscr{B}_1$, which contains the only possible Gram matrix in one dimension, and then repeatedly applied until we obtain $\mathscr{B}_9$. Since we start with only finitely many lattices, we only produce finitely many new lattices in each step and hence $\mathscr{B}_{n}$ is finite for all $n$. 

In this form, the algorithm also produces non-primitive lattices. However, it can be shown that the non-primitive cases are excluded if we consider only non-vanishing $v$ in each step, and that we do not accidentally remove some primitive lattices this way. At this point it should also be noted that, because this naive algorithm very slow for large $n$, the actual implementation includes several optimizations. 

Finally we have to check for the existence of pathological cases. This is done as follows: for each lattice we found (including the non-primitive ones obtained by considering orthogonal sums), one constructs all overlattices $\Xi^\prime$ that are obtained by including additional norm 3 vectors. If we encounter a primitive lattice that was not yet obtained, we must add it to our results.

The actual computation showed that for $n \leq 9$ no pathological case exists. Also, in Table~\ref{t:xi} the number of primitive basic lattices that were obtained for each dimension is listed.

\begin{table}
\centering
\begin{tabular}{crrrr}
\toprule
$\dim(\Xi)$ & PB & PB + (\ref{maineq1}) & PBS & PE\\
\midrule
1 & 1 & 1 & 1 & 1 \\
2 & 2 & 0 & 0 & 0 \\
3 & 7 & 1 & 1 & 1 \\
4 & 28 & 0 & 0 & 0 \\
5 & 136 & 1 & 1 & 1 \\
6 & 911 & 6 & 6 & 3 \\
7 & 8665 & 2 & 2 & 2 \\
8 & 131316 & 8 & 7 & 4 \\
9 & 3345309 & 40 & 36 & 14 \\
\midrule
Total & 3486375 & 59 & 54 & 26 \\
\bottomrule
\end{tabular}
\caption{The number of lattices with the respective properties: \textit{basic} (B), \textit{supersymmetry} (S), \textit{primitivity} (P), \textit{elementarity} (E).}
\label{t:xi}
\end{table}

\subsubsection*{The Primitive Elementary Lattices}
At this stage we possess a list of all lattices with the basic and primitivity properties. The next step is to solve the conditions imposed by world-sheet supersymmetry. First, it is practical to consider only equation~(\ref{maineq1}), which is a system of linear equations in the $\vert A(s) \vert^2$. It turns out that in total only 59 lattices possess a solution, so the number of candidate lattices is drastically reduced (cf. Table~\ref{t:xi}). For these candidates we then consider, in a somewhat case by case manner, the full set of equations~(\ref{maineq1})--(\ref{maineq3}). This is done using the following strategies:
\begin{itemize}
\item By means of the $U(1)^9$ symmetry discussed in Subsection~\ref{ss:chiral}, we may fix the phases of some $A(s)$. Moreover, if $U(1)^9$ is extended to some non-Abelian symmetry, we may use this symmetry to set some $A(s)$ to zero. This procedure radically reduces the complexity of the problem in most cases.
\item Sometimes, systems of polynomial equations are easier to solve if one first computes a Gr\"obner basis of the corresponding ideal. Especially, if the computed Gr\"obner basis is trivial then the system has no solutions. 
\end{itemize}
With these methods it was possible to rule out the existence of a solution for $5$ out of the $59$ candidate lattices. For further $26$ candidates an explicit solution was found, albeit with some trial and error. Then, it was proven that these lattices are elementary and that all remaining candidates can be reproduced by gluing together orthogonal sums thereof. 

Thus, for $\dim(\Xi) \leq 9$, there exist $54$ primitive admissible lattices of which $26$ are primitive elementary. By orthogonally combining them one obtains in $\dim(\Xi) = 9$ a total of $63$ admissible lattices, and $32$ of them are elementary. A summary of these results is shown in Table~\ref{t:xi}. 

In the rest of the paper we identify a primitive elementary lattice by its dimension, and, in the cases $6 \leq \dim(\Xi) \leq 9$, also by an additional uppercase Latin subscript which is assigned in alphabetical order. Moreover, we use a shorthand notation where e.g. $3^16_A^1$ denotes the orthogonal sum of the primitive elementary lattices $3$ and $6_A$. Gram matrices for the $26$ primitive elementary lattices are provided in Table~\ref{t:gram}, together with some further information.

\subsubsection*{The Lattice Inclusion Graph}
For each of the 32 elementary supercurrent lattices that we classified one now constructs a right-mover lattice $(\Gamma_{14})_\text{R}$ from the constraint vectors $(0; s, v)$. These right-mover lattices are listed in Table~\ref{t:bottoms}. They are minimal, in the sense that there is no solution to the supersymmetry constraints~(\ref{maineq1})--(\ref{maineq3}) for any strictly smaller sublattice. However, any even overlattice
\begin{align}
(\Gamma_{14})_\text{R}^\prime \supseteq (\Gamma_{14})_\text{R}
\end{align}
clearly inherits the solution $A(s)$ from $(\Gamma_{14})_\text{R}$ (it may violate the chiralness constraint from Subsection~\ref{ss:chiral}, though), and only finitely many such overlattices can exist. The explicit construction of all these overlattices produced a total of 414 right-mover lattices.

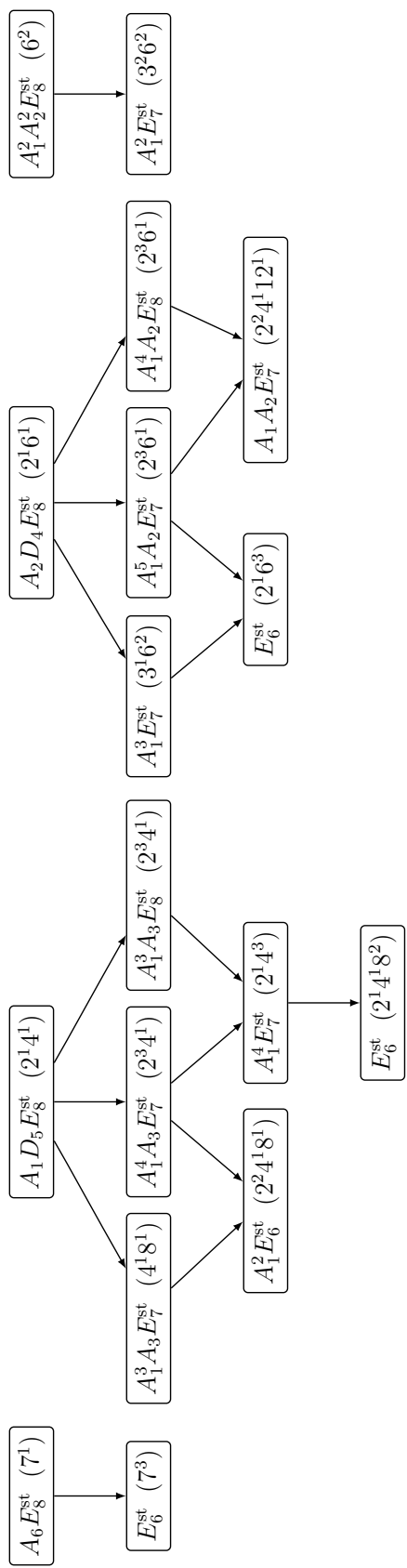
\begin{sidewaysfigure}[p]
\centering
\adjustbox{scale = 0.87}{\adjustbox{valign=t}{\input{C9.tex}}\;\;\adjustbox{valign=t}{\input{C7.tex}}\;\;\adjustbox{valign=t}{\input{C3.tex}}\;\;\adjustbox{valign=t}{\input{C2.tex}}}
\caption{Lattices $(\Gamma_{14})_\text{R}$ in $\mathscr{G}_3$, $\mathscr{G}_4$, $\mathscr{G}_5$ and $\mathscr{G}_9$ (from left to right) with spacetime supersymmetry. The node label indicates the root system of norm 2 vectors and the elementary divisors of $(\Gamma_{14})_\text{R}$.}
\label{f:C3459}
\end{sidewaysfigure}

From these lattices it is possible to construct a directed graph $\mathscr{G}$ in which each lattice is represented by a node, and two lattices $A$ and $B$ are connected by an arrow, $A \longrightarrow B$, if $A \supset B$ and the index $\vert A / B \vert$ is prime. It turns out that the graph splits into nine disjoint connected components $\mathscr{G}_1$ to $\mathscr{G}_9$. Consequently, for most right-mover lattices, more than one of the 32 elementary supercurrents can be chosen, and, as pointed out in Subsection~\ref{ss:chiral}, it is possible that these choices are related by a symmetry transformation. In Figures~\ref{f:C1}--\ref{f:C3459} we display the connected components $\mathscr{G}_1$ to $\mathscr{G}_5$ and $\mathscr{G}_9$, but only the subgraphs thereof consisting of nodes with spacetime supersymmetry. Also, Table~\ref{t:graphstat} lists the number of lattices in each connected component, separately for the different levels of spacetime supersymmetry. It turns out that among all 414 lattices, only 99 comply with the chiralness condition introduced in~Subsection~\ref{ss:chiral}, and merely 19 lead to $\mathcal{N}=1$ spacetime supersymmetry. These 19 lattices and some of their properties are listed in Table~\ref{t:susy}. 

\begin{table}
\centering
\begin{tabular}{crrrrrr}
\toprule
  &  $D_5^\text{st}$ &  $E_6^\text{st}$ &  $E_7^\text{st}$ &  $E_8^\text{st}$ & $D_{5+k}^\text{st}$ & Total\\
\midrule
$\mathscr{G}_1$   &   19 &    6 &    6 &    3 &   36 &   70 \\
$\mathscr{G}_2$    &   35 &    9 &    9 &    4 &  115 &  172 \\
$\mathscr{G}_3$    &    1 &    1 &    0 &    1 &    1 &    4 \\
$\mathscr{G}_4$    &    8 &    2 &    3 &    2 &   45 &   60 \\
$\mathscr{G}_5$    &   10 &    1 &    3 &    2 &   41 &   57 \\
$\mathscr{G}_6$    &    2 &    0 &    0 &    1 &    5 &    8 \\
$\mathscr{G}_7$    &    1 &    0 &    0 &    1 &    6 &    8 \\
$\mathscr{G}_8$    &    1 &    0 &    0 &    1 &   10 &   12 \\
$\mathscr{G}_9$   &    3 &    0 &    1 &    1 &   18 &   23 \\
\midrule
Total $\mathscr{G}$ & 80 & 19 & 22 & 16 & 277 & 414 \\
\bottomrule
\end{tabular}
\caption{Statistics of the lattice inclusion graph $\mathscr{G}$ in total, and also separately for each connected component $\mathscr{G}_i$. Denoted are the total number of lattices, as well as the number of lattices with certain spacetime sublattice ($k > 0$).}
\label{t:graphstat}
\end{table}

It is now worth to discuss the following special nodes:
\begin{enumerate}
\item A bottom node is a node $B$ for which there do not exist other nodes $B^\prime \subset B$.
\item A top node is a node $T$ for which there do not exist other nodes $T^\prime \supset T$.
\end{enumerate} 
The bottom nodes in our graph are clearly those representing the 32 lattices $(\Gamma_{14})_\text{R}$ constructed from the elementary supercurrent lattices $\Xi$. The top nodes of our graph are listed in Table~\ref{t:tops}. Remarkably, each connected component contains, among others, a top node representing a $\mathcal{N}=4$ theory. These theories are Narain-compactifications of the ten-dimensional theory where $(\Gamma_{14})_\text{R}$ is of the form $(\Gamma_{6})_\text{R} \oplus E_8^\text{st}$ and $(\Gamma_{6})_\text{R}$ is the root lattice of a rank 6 semi-simple Lie algebra of ADE type.


\subsection{Classification of Left-Mover Lattices}\label{ss:LM}
In the last subsection we classified the possible right-mover lattices $(\Gamma_{14})_\text{R}$ that can appear in a chiral four-dimensional covariant lattice model. We will now discuss the corresponding left-mover lattices.

\subsubsection*{Discriminant Forms}
It is first necessary to introduce the concept of discriminant forms~\cite{Nikulin:1980}. Let $\Lambda$ be a lattice with Gram matrix $G$, and let $\det(\Lambda) = \vert\text{det}(G)\vert$ denote its determinant. Then, for any sublattice $\Lambda^\prime \subseteq \Lambda$ one has $\vert\Lambda/\Lambda^\prime\vert^2 = \det(\Lambda^\prime)/\det(\Lambda)$. Let further $\Lambda^\ast$ denote the dual lattice, which is defined to be the lattice of all vectors in the $\mathbb{R}$-span of $\Lambda$ that have integral inner product with all vectors of $\Lambda$. Clearly, if $\Lambda$ is integral then $\Lambda \subseteq \Lambda^\ast$, so we can define the quotient group $\Lambda^\ast/\Lambda$. This quotient is a product of (finite) cyclic groups whose orders are given by the elementary divisors of $\Lambda$. These are in turn defined to be the elementary divisors obtained from the Smith normal form of $G$. One further sees that $\vert\Lambda^\ast/\Lambda \vert = \det(\Lambda)$.

Let us now assume that $\Lambda$ is an even lattice. Then, we can introduce a quadratic form $Q_\Lambda : \Lambda^\ast/\Lambda \mapsto \mathbb{Q}/2\mathbb{Z}$, given by
\begin{align}
Q_\Lambda(v + \Lambda) = v^2\,,
\end{align}
for $v \in \Lambda^\ast$. This is indeed well defined because $(v + x)^2 - v^2 \in 2\mathbb{Z}$ for all $x\in\Lambda$. The quotient $\Lambda^\ast/\Lambda$ together with the quadratic form $Q_\Lambda$ is called discriminant form $\text{disc}(\Lambda)$ of $\Lambda$. In particular, for a self-dual lattice $\Lambda = \Lambda^\ast$, and hence $\text{disc}(\Lambda)$ is trivial. Moreover, we define an isomorphism between discriminant forms, $\phi : \text{disc}(\Lambda_1) \mapsto \text{disc}(\Lambda_2)$, to be a group isomorphism that also preserves the quadratic form, i.e. $Q_{\Lambda_2} \circ \phi = Q_{\Lambda_1}$.

Now, there is a well known theorem~\cite{SPLAG}: for any self-glue free decomposition of the form $\Lambda \supset \Lambda_1 \oplus \overline{\Lambda}_2$ with $\Lambda$ even and self-dual, one obtains an isomorphy
\begin{align}\label{eqiso}
	\text{disc}(\Lambda_1) \cong \text{disc}(\Lambda_2).
\end{align} 
Here, $\dim(\Lambda) = \dim(\Lambda_1) + \dim(\Lambda_2)$ is assumed. Let us now prove this isomorphy by simple means. First, one realizes that each coset in $\Lambda/(\Lambda_1 \oplus \overline{\Lambda}_2)$ can be uniquely represented by a pair $(\xi_1,\xi_2)$ of cosets $\xi_1 \in \Lambda_1^\ast/\Lambda_1$ and $\xi_2\in\Lambda_2^\ast/\Lambda_2$. Since we are considering a self-glue free decomposition, each coset in $\Lambda_1^\ast/\Lambda_1$ and $\Lambda_2^\ast/\Lambda_2$ can appear \textit{at most once} in such a pair (assuming otherwise immediately produces a contradiction). This gives us the following inequalities:
\begin{align}\label{ineq12}
\left\vert \frac{\Lambda}{\Lambda_1 \oplus \overline{\Lambda}_2} \right\vert \leq \left\vert \frac{\Lambda_i^\ast}{\Lambda_i} \right\vert\,, \qquad i\in\lbrace1,2\rbrace\,.
\end{align}
Also, since $\Lambda$ is self-dual, each $\xi_1 \in \Lambda_1^\ast/\Lambda_1$ and $\xi_2 \in \Lambda_2^\ast/\Lambda_2$ must appear \textit{at least once} in a pair $(\xi_1,\xi_2)$. This can be seen by deriving
\begin{align}
\left\vert \frac{\Lambda}{\Lambda_1 \oplus \overline{\Lambda}_2} \right\vert^2 = \frac{\det (\Lambda_1 \oplus \overline{\Lambda}_2)}{\det(\Lambda)} = \left\vert \frac{\Lambda_1^\ast}{\Lambda_1} \right\vert \left\vert \frac{\Lambda_2^\ast}{\Lambda_2} \right\vert \,,
\end{align}
using $\det(\Lambda) = 1$. Hence, in~(\ref{ineq12}) equalities must hold and the set of pairs $(\xi_1, \xi_2)$ defines a bijective map $\phi: \Lambda_1^\ast/\Lambda_1 \mapsto \Lambda_2^\ast/\Lambda_2$ which preserves the group structure. Thus, we have established the following group isomorphies: 
\begin{align}
	\frac{\Lambda}{\Lambda_1 \oplus \overline{\Lambda}_2} \cong \frac{\Lambda_1^\ast}{\Lambda_1} \cong \frac{\Lambda_2^\ast}{\Lambda_2}
\end{align}
Furthermore, from the fact that $\Lambda$ is even it follows that $Q_{\Lambda_1}(\xi_1) - Q_{\Lambda_2}(\phi(\xi_1)) = 0$, which proves the theorem. Here, the minus sign originates from the fact that $\Lambda_2$ appears with negated inner product in the decomposition of $\Lambda$.

It is also possible to show the following ``converse" statement: given two even lattices $\Lambda_1$, $\Lambda_2$ and an isomorphism $\phi: \text{disc}(\Lambda_1) \mapsto \text{disc}(\Lambda_2)$, one can construct a lattice
\begin{align}
\Lambda = \bigcup_{\xi\in\text{disc}(\Lambda_1)} \xi \times \phi(\xi)\,,
\end{align}
which is self-dual and decomposes as $\Lambda \supset \Lambda_1 \oplus \overline{\Lambda}_2$. Here, the cartesian product $\xi \times \phi(\xi)$ is to be interpreted as an element of $(\Lambda_1 \oplus \overline{\Lambda}_2)^\ast / (\Lambda_1 \oplus \overline{\Lambda}_2)$. 

\subsubsection*{Lattice Genera and the Mass Formula}
Now, we want to introduce the concept of lattice genera. Let $\Lambda_1$ and $\Lambda_2$ denote two integral lattices with Gram matrices $G_1$ and $G_2$. Then one defines an equivalence relation ``$\equiv$" as follows: we say that $\Lambda_1 \equiv \Lambda_2$ if for every prime number $p$ there exists an invertible $p$-adic integral matrix $U_p$ such that
\begin{align}
	U_p G_1 U_p^T = G_2\,,
\end{align}
and if further $\Lambda_1$ and $\Lambda_2$ have the same signature. The corresponding equivalence classes are called genera. An alternative characterization of the genus is due to Nikulin~\cite{Nikulin:1980}: two even lattices $\Lambda_1$ and $\Lambda_2$ lie in the same genus $\mathcal{G}$ if and only if they have identical signature and their discriminant forms are isomorphic (the isomorphy already implies $p_1 - q_1 = p_2 - q_2 = 0 \text{ mod } 8$ for the respective signatures $(p_1, q_1)$ and $(p_2, q_2)$). In particular, two lattices in the same genus have identical elementary divisors.

There is a classic result that states that a genus $\mathcal{G}$ contains only finitely many lattices, and the predominant method for the enumeration of all lattices in a genus is known as Kneser's neighborhood method~\cite{Kneser:1957}. This method is related to the ``shift vector method" in Appendix A.4 of~\cite{Lerche:1988np}, which is in turn related to certain shift-orbifold constructions. Also, in some cases the ``replacement" lattice engineering method (cf. Appendix A.4 of~\cite{Lerche:1988np}) turns out to be useful. 

Another relevant tool in the study of lattice genera is the Smith-Minkowski-Siegel mass formula~\cite{Siegel:1935, Conway:1988}. The mass of a genus is defined as
\begin{align}\label{sms}
	m(\mathcal{G}) = \sum_{\Lambda \in \mathcal{G}} \frac{1}{\vert \text{Aut}(\Lambda) \vert}\,.
\end{align}
Here, $\text{Aut}(\Lambda)$ denotes the automorphism group (point group) of $\Lambda$ (definiteness of the lattices is assumed). Note, that the definition of the mass depends on \textit{all} lattices in $\mathcal{G}$. The mass formula then provides another way of computing the mass which only requires explicit knowledge of a \textit{single} $\Lambda \in \mathcal{G}$. This computation is rather complicated (the technicalities are found in~\cite{Conway:1988}) and we will not go into the details here. 

An important application of the mass formula is the computation of a lower bound on $|\mathcal{G}|$: from $\vert \text{Aut}(\Lambda) \vert \geq 2$ one obtains
\begin{align}
	\vert \mathcal{G} \vert\geq 2 m(\mathcal{G})\,.
\end{align}
However, this bound is rather crude in many cases. The mass formula also allows to verify whether an explicit enumeration of lattices in a genus is exhaustive. 

\subsubsection*{The Genera of Left-Mover Lattices}
Let us now study the left-mover lattices using the framework of lattice genera and discriminant forms.

First, using the theorem proved earlier, we conclude from the self-duality of $\Gamma_{22,14}$ that the lattices $(\Gamma_{22})_\text{L}$ and $(\Gamma_{14})_\text{R}$ appearing in the decomposition~(\ref{chiraldecomp}) must have isomorphic discriminant forms. Hence, by the theorem of Nikulin~\cite{Nikulin:1980}, the right-mover lattice completely determines the genus $\mathcal{G}_\text{L}$ of $(\Gamma_{22})_\text{L}$. Moreover, $(\Gamma_{22})_\text{L}$ can be replaced by any other lattice from $\mathcal{G}_\text{L}$ without destroying self-duality, so set of possible left-mover lattices that can be paired with some specific $(\Gamma_{14})_\text{R}$ is given precisely by the corresponding genus $\mathcal{G}_\text{L}$. Not surprisingly, it is also possible to exchange $(\Gamma_{14})_\text{R}$ with a different lattice from the same genus, provided that it also obeys the constraints from world-sheet supersymmetry. 

For some of the right-mover lattices belonging to the lattice inclusion graph $\mathscr{G}$ that we constructed in Subsection~\ref{ss:RM}, a computational analysis of the corresponding genera $\mathcal{G}_\text{L}$ was performed. First, we consider the left-mover lattices corresponding to the top nodes. Their special importance is that they may serve as a starting point for the enumeration of all the other relevant genera. Remarkably it turns out that, separately for each connected component of $\mathscr{G}$, the respective top nodes belong to the same genus. Then, by means of the ``replacement" method described in Appendix A.4 of~\cite{Lerche:1988np}, it was possible to completely classify these genera (in some cases, for practicability reasons a generalized method which involves also odd lattices was used). In Table~\ref{t:tops}, the respective $\vert \mathcal{G}_\text{L}\vert$ are listed. The genus $\mathcal{G}_\text{L}$ corresponding to the top nodes of $\mathscr{G}_1$ was, using the same method, already classified in~\cite{Beye:2013moa}.  

For the bottom nodes of $\mathscr{G}$, a lower bound on the respective $\vert \mathcal{G}_\text{L} \vert$ was calculated using the Smith-Minkowski-Siegel mass formula. In order to apply this formula, we need a representative of each $\mathcal{G}_\text{L}$. Such a representative is given e.g. by $(\Gamma_{14})_\text{R} \oplus E_8$. The results are displayed in Table~\ref{t:bottoms}. One recognizes that from the bottom nodes alone one obtains a total of at least $O(10^{23})$ models. Of course, these models are not guaranteed to be chiral. They are also not supersymmetric, so supposedly many of them contain tachyons. However, in any case these high numbers rule out an explicit enumeration and evaluation of all these models. 

Finally, let us consider the 19 right-mover lattices in Table~\ref{t:susy} that lead to $\mathcal{N}=1$ supersymmetry. First, for the lattice $A_2^4 E_6^\text{st}\;(3^3)$ which is contained in $\mathscr{G}_1$ (cf. Figure~\ref{f:C1}), a complete classification of the corresponding genus $\mathcal{G}_\text{L}$ was performed by considering certain kinds of shift-orbifolds (see also Subsection~\ref{ss:asymm}). It was found that $\vert\mathcal{G}_\text{L}\vert = 2030$, so one can construct $2030$ models from this right-mover lattice (in this case, there is only one inequivalent embedding of the form~(\ref{chiraldecomp})). Interestingly, the right-mover lattices $A_2^4 E_6^\text{st}\;(3^3)$ and $A_2^3 E_8^\text{st}\;(3^3)$ belong to the same genus. Hence, the $2030$ left-mover lattices we classified also appear in Narain-compactified $\mathcal{N}=4$ models. For the other lattices in Table~\ref{t:susy}, only a lower bound on the respective $\vert\mathcal{G}_\text{L}\vert$ was calculated using the mass formula (in the case of the lattice with elementary divisors $2^6$ the resulting lower bound was less than one and therefore meaningless).

The bottom line is that, even in the $\mathcal{N}=1$ case, we expect at least $O(10^{10})$ models. However one must keep in mind the crudeness of the lower bound. Out of curiosity, we can quantify this crudeness in the case of the right-mover lattice $A_2^4 E_6^\text{st}\;(3^3)$ where we enumerated $\mathcal{G}_\text{L}$ exactly: there, the lower bound calculated using the mass formula is of $O(10^{-3})$, thus the deviation is of $O(10^6)$.


\subsection{Relation to Asymmetric Orbifolds}\label{ss:asymm}
In~\cite{AsymOrbCL}, an equivalence between certain asymmetric orbifolds and covariant lattice models was found. Here, we want to discuss this equivalence in the light of our results.

\subsubsection*{The Asymmetric Orbifold Construction}
Let us briefly introduce the $Z_N$ asymmetric orbifold construction from~\cite{AsymOrbCL}. There, one starts with a Narain-compactified $\mathcal{N}=4$ theory and twists the six compactified right-moving bosons $X^i$ and fermions $\psi^i$ as
\begin{align}
	X^i &\mapsto {\theta^i}_j X^j \\
	\psi^i &\mapsto {\theta^i}_j \psi^j  \,,
\end{align}
where $\theta$ is assumed to be non-degenerate, i.e. $\det (1 - \theta) \neq 0$. This twisting preserves the world-sheet supercurrent,
\begin{align}
G(z) &= i\psi(z) \cdot \partial X(z)\,.
\end{align}
The left-movers may also be subject to a shift action with shift vector $v_\text{L}$. By bosonizing the $\psi^i$, one sees that the original $\mathcal{N}=4$ theory is equivalent to a covariant lattice $\Gamma_{22,14}$ which contains an $\overline{E_8^\text{st}}$ sublattice. In this bosonized description, the twist action on the fermions $\psi^i$ is replaced by a shift action with shift vector $v_\psi$. The main result of of~\cite{AsymOrbCL} is that, under certain circumstances, the twist action on the $X^i$ can, by a change of basis, also be turned into a shift action with some shift vector $v_X$. This change of basis is only possible if the Narain lattice decomposes as 
\begin{align}
	\Gamma_{22,6} \supset (\Gamma_{22})_\text{L} \oplus \overline{(\Gamma_{6})}_\text{R} \,,
\end{align}
and if $(\Gamma_{6})_\text{R}$ is the root lattice of a rank 6 semi-simple Lie algebra of ADE type. Furthermore, the twist must be an element of the Weyl-group of this Lie algebra. Then, $\Gamma_{22,14}$ decomposes as in equation~(\ref{chiraldecomp}) with 
\begin{align}
(\Gamma_{14})_\text{R} = (\Gamma_{6})_\text{R} \oplus E_8^\text{st}\,,
\end{align}
and, in the new basis, the supercurrent takes the form of equation~(\ref{ansatz}). Hence, it is possible to identify $(\Gamma_{14})_\text{R}$ in our lattice inclusion graph $\mathscr{G}$. In~\cite{AsymOrbCL}, all choices of $\theta$ along with the corresponding shift vectors $v_X$ and $v_\psi$ were classified for each $(\Gamma_{6})_\text{R}$.

Let us now discuss the orbifold theory that is obtained from the complete shift action. In the notation introduced in Subsection~\ref{ss:covlat}, the shift vector can be compactly written as $v = (v_\text{L}; v_\text{R}, 0)$, where $v_\text{R}$ is composed of $v_X$ and $v_\psi$. Also, if the twist-orbifolding defines a modular invariant theory, one can always choose $v$ such that $v^2 \in 2\mathbb{Z}$ (cf. Appendix A.4 of~\cite{Lerche:1988np}). Let in the following $N$ denote the smallest natural number such that $Nv \in \Gamma_{22,14}$ (because we were assuming a nontrivial twist $\theta$, one can show that $N > 1$). 

The untwisted sector of the orbifold theory is then represented by the sublattice $\Gamma_{22,14}^\text{u}$ of vectors $x \in \Gamma_{22,14}$ with vanishing orbifold phase, i.e. $x \cdot v \in \mathbb{Z}$. This lattice always has a (self-glue free) decomposition
\begin{align}
	\Gamma_{22,14}^\text{u} \supset (\Gamma_{22}^\text{u})_\text{L} \oplus \overline{(\Gamma_{14}^\text{u})}_\text{R} \,.
\end{align}
Note that because the original twisting preserves the world-sheet supercurrent, the supercurrent of the form~(\ref{ansatz}) obtained on $(\Gamma_{14})_\text{R}$ can also be used on $(\Gamma_{14}^\text{u})_\text{R}$. Thus, $(\Gamma_{14}^\text{u})_\text{R}$ must be an element of $\mathscr{G}$. Moreover, $\vert \Gamma_{22,14} /\Gamma_{22,14}^\text{u}\vert = N$, so $\Gamma_{22,14}^\text{u}$ is not self-dual. However, the inclusion of the twisted sectors provides additional glue vectors of the form
\begin{align}
nv + \Gamma_{22,14}^\text{u}\,, \qquad n \in \lbrace 1, \ldots, N-1\rbrace\,.
\end{align}
These cosets complete $\Gamma_{22,14}^\text{u}$ to an even self-dual lattice $\Gamma_{22,14}^\prime \supset \Gamma_{22,14}^\text{u}$. Then, in particular $(\Gamma_{14}^\prime)_\text{R} \supseteq (\Gamma_{14}^\text{u})_\text{R}$, where the equality is achieved for appropriate choices of $v_\text{L}$. Note that the twisted sectors may also increase the amount of spacetime supersymmetry. 

\subsubsection*{The Right-Mover Lattices of Twist-Orbifolds}
Now, it is interesting to investigate what lattices $(\Gamma_{14}^\text{u})_\text{R}$ actually arise due to this mechanism. Let us restrict to those $(\Gamma_{6})_\text{R}$ and twists $\theta$ that were found in~\cite{AsymOrbCL} to lead to $\mathcal{N}=1$ spacetime supersymmetry (cf. the summary in Table~\ref{t:orb}). Then, the obtained right-mover lattices $(\Gamma_{14}^\text{u})_\text{R}$ must be among those 19 right-mover lattices in Table~\ref{t:susy}. An explicit calculation verified that this in fact happens. Also, in Table~\ref{t:susy} it is denoted which type of orbifold corresponds to which right-mover lattice. Interestingly, in several cases it happens that different types of twist-orbifold lead to the same right-mover lattice. One such case is given by the $Z_6^\text{I}$ and $Z_6^\text{II}$ orbifolds constructed from $(\Gamma_{6})_\text{R} = E_6$, as both lead to the $A_1^4E_6^\text{st}\;(3^16^2)$ right-mover lattice shown in Figure~\ref{f:C1}. This could (but not necessarily must) mean that some models can be obtained from either twist-orbifold construction. 

Let us finally treat the question whether all elementary supercurrents can be obtained from the construction of~\cite{AsymOrbCL} (or a possible generalization thereof that also includes e.g. the $Z_N \times Z_M$ case). In~\cite{AsymOrbCL}, a simple condition is provided that allows to check whether a given admissible $\Xi$ can be obtained from a twist-orbifold construction: there must exist three orthonormal vectors $e^I$ in $\Xi^\ast$, i.e. $e^I \cdot e^J = \delta^{IJ}$, so that for each norm 3 vector $t \in \Xi$ there is exactly one $e^I$ such that $t \cdot e^I = \pm 1$.

We explicitly checked this condition for all $32$ elementary lattices $\Xi$ that resulted from our classification. It turned out that, except for cases $1^27_A^1$, $1^18_A^1$ and $9_A^1$, it was possible to satisfy the condition. Moreover, for these exceptional cases the corresponding $(\Gamma_{14})_\text{R}$ are not contained in a lattice of the form $(\Gamma_{6})_\text{R} \oplus E_8^\text{st}$ (note, that this would be required if above condition were true). Hence, it is impossible for them to appear in a $\mathcal{N} = 1$ model.

\begin{table}
\centering
\begin{tabular}{cllllll}
\toprule
Component & $(\Gamma_6)_\text{R}$ & Type & Carter Diagram &  $v_\psi$ & $v_X$ \\
\midrule
\multirow{7}{*}{$\mathscr{G}_1$} & \multirow{4}{*}{$E_6$} & $Z_3$ & $A_2^3$ & $(1, 1, -2)/3$ &  $(1,1,0,1,1,1)/3$\\
				& & $Z_6^\text{I}$ & $E_6(a_2)$ & $(1,1,-2)/6$ & $(1,0,1,0,1,0)/6$\\
				& & $Z_6^\text{II}$ & $A_1A_5$ & $(1,2,-3)/6$ & $(1,1,1,1,1,3)/6$\\
				& & $Z_{12}^\text{I}$ & $E_6$ & $(1,4,-5)/12$ & $(1,1,1,1,1,1)/12$\\
	\cmidrule(l){2-6}
& $A_1A_5$ & $Z_6^\text{II}$ & $A_1A_5$ & $(1,2,-3)/6$ & $(3,1,1,1,1,1)/6$ \\
	\cmidrule(l){2-6}
& $A_2^3$ & $Z_3$ & $A_2^3$ & $(1, 1, -2)/3$ &  $(1,1,1,1,1,1)/3$\\
\midrule
\multirow{5}{*}{$\mathscr{G}_2$} & \multirow{2}{*}{$D_6$} & $Z_4$ & $A_1^2D_4{(a_1)}$ or $A_3^2$ & $(1,1,-2)/4$ & $( 2, 0, 1, 0, 1, 1 )/4$\\
				& & $Z_8^\text{I}$ & $D_6(a_1)$ & $(1,2,-3)/8$ & $(1,1,0,1,1,1)/8$ \\
	\cmidrule(l){2-6}
& $A_3^2$ & $Z_4$ & $A_3^2$ & $(1,1,-2)/4$ & $(1,1,1,1,1,1)/4$ \\
	\cmidrule(l){2-6}
& $A_1^2D_4$ & $Z_4$ & $A_1^2D_4(a_1)$ & $(1,1,-2)/4$ & $(2,2,1,0,1,1)/4$\\
\midrule
$\mathscr{G}_3$ & $A_6$ & $Z_7$ & $A_6$ & $(1,2,-3)/7$ & $(1,1,1,1,1,1)/7$ \\
\midrule
$\mathscr{G}_4$ & $A_1D_5$ & $Z_8^\text{II}$ & $A_1D_5$ & $(1,3,-4)/8$ & $(4,1,1,1,1,1)/8$ \\
\midrule
$\mathscr{G}_5$ & $A_2D_4$ & $Z_6^\text{II}$ & $A_2D_4$ & $(1,2,-3)/6$ & $(2,2,1,1,1,1)/6$ \\
\bottomrule
\end{tabular}
\caption{All different types of twist-orbifolds that were found in~\protect\cite{AsymOrbCL} to lead to $\mathcal{N}=1$ covariant lattice theories. For the $v_X$, the notation of~\protect\cite{AsymOrbCL} is adopted.}
\label{t:orb}
\end{table}

\subsubsection*{A Class of $Z_3$ Asymmetric Orbifold Models}
In order to construct a complete asymmetric orbifold model, one must also provide a left-mover lattice $(\Gamma_{22})_\text{L}$ and a shift vector $v_\text{L}$ as input. Then, one can calculate the lattices $(\Gamma_{22}^\prime)_\text{L}$ and $(\Gamma_{14}^\prime)_\text{R}$, as well as their embedding into $\Gamma^\prime_{22,14}$. In fact one can show that, by considering all lattices in $(\Gamma_{22})_\text{L} \in \mathcal{G}_\text{L}$ and all modular invariant shift vectors, it is possible to completely classify the genera $\mathcal{G}_\text{L}^\prime$ of lattices $(\Gamma_{22}^\prime)_\text{L}$. 

This was done explicitly in the particular case of the $Z_3$-orbifold with $(\Gamma_6)_\text{R} = E_6$, where we also require that $v_\text{L}$ is of order 3. In that case, $\mathcal{G}_\text{L}$ consists of 31 lattices and a modular invariant shift vector satisfies
\begin{align}
	3 v_\text{L} &\in (\Gamma_{22})_\text{L}\\
	v_\text{L} &\notin (\Gamma_{22})_\text{L}^\ast\\
	3 v_\text{L}^2 &\in 2 \mathbb{Z}\,.
\end{align} 
One also sees that this implies $(\Gamma_{22}^\prime)_\text{L} = (\Gamma_{22}^\text{u})_\text{L}$ and $(\Gamma_{14}^\prime)_\text{R} = (\Gamma_{14}^\text{u})_\text{R}$. Furthermore, two modular invariant shift vectors $v_\text{L}$ and $v_\text{L}^\prime$ produce identical $(\Gamma_{22}^\prime)_\text{L}$ if 
\begin{align}
	v_\text{L}^\prime - \theta v_\text{L} \in (\Gamma_{22})_\text{L}^\ast
\end{align}
for some automorphism $\theta$ of $(\Gamma_{22})_\text{L}$. This fact can be used to reduce the number of shift vectors one has to check for each left-mover lattice in $\mathcal{G}_\text{L}$ to a finite and tractable number. 

Then, by carrying out the shift-orbifolding procedure in each case, a total of $2030$ inequivalent lattices $(\Gamma_{22}^\prime)_\text{L}$ were obtained and it was verified that they constitute the left-mover genus $\mathcal{G}^\prime_\text{L}$ corresponding to $(\Gamma_{14}^\prime)_\text{R}$. The phenomenology of two of these asymmetric orbifold models, and also of another $Z_3$ model from $(\Gamma_6)_\text{R} = A_2^3$, was already studied in detail in~\cite{Beye:2013ola}.


\section{Conclusions and Outlook}
In this work, chiral four-dimensional covariant lattice models were revisited and a classification of all possible right-mover lattices was performed. The result is that there are in total 99 right-mover lattices which may lead to chiral models, and only 19 of them lead to $\mathcal{N}=1$ spacetime supersymmetry. Also, it was found that once a right-mover lattice is fixed, modular invariance requires that the set of possible left-mover lattices forms a genus. Then, some of the relevant genera were either enumerated completely, or a lower bound on their order was given using the Smith-Minkowski-Siegel mass formula. Finally, we studied how the equivalence between certain covariant lattice and twist-orbifold models fits into our picture, and found that there exist some covariant lattices which cannot be obtained as a twist-orbifold theory.

Especially the 19 right-mover lattices that lead to $\mathcal{N}=1$ spacetime supersymmetry might be interesting for model building. Some models based on these lattices were already considered explicitly as an asymmetric orbifold, but there are still at least $O(10^{10})$ models lying around to be studied. As in the case of the genus corresponding to $E_6/Z_3$ orbifold models that we enumerated completely, some smaller genera (e.g. the $D_6/Z_4$ case) may be studied exactly. However, a complete evaluation of the larger genera does not seem to be practicable, both from the viewpoints of computation time and required memory. Nevertheless, one might resort to other methods. For example, a randomized search that just produces a large number of models is perfectly viable, as long as one does not mind obtaining duplicate models. Another approach would be to impose more phenomenological constraints, e.g. one could require that the left-mover lattice contains an $A_1A_2$ factor corresponding to $SU(2)_L \times SU(3)_C$. Then, it might be possible to circumvent the lower bounds that we calculated.

There is another remark on the supercurrent lattices $\Xi$ that we found. Here, we used a rather brute force approach to classify them. However, it would be interesting to have a more fundamental and geometrical understanding of these lattices, maybe in a way similar to how we understand root systems in terms of simple roots. One could also ask which of the admissible lattices $\Xi$ allow for an additional world-sheet supercurrent that completes the $N = 2$ super-Virasoro algebra. Clearly, such a supercurrent must be allowed for the admissible lattices obtained by extracting the norm 3 vectors $s$ from the constraint vectors $(0;s,v)$ of a right-mover lattice with $\mathcal{N}=1$ spacetime supersymmetry. An example for this would be the $9_N$ supercurrent lattice that appears in the $A_6/Z_7$ orbifold. Also, the one-dimensional supercurrent lattice allows for an additional world-sheet supersymmetry because it corresponds to a $N = 2$ minimal model.

Furthermore, the lattice theories discussed here only cover CFTs with Kac-Moody algebras of level one, so a generalization that covers also higher levels would be desirable (note that a generalization of the theory of lattice genera to general CFTs was attempted in \cite{Hohn:2003}). Nevertheless, our results may be useful in the construction of some sort of hybrid models. For example, one could combine our primitive elementary lattices with $N = 2$ minimal models to fill up the required central charge $c_\text{R} = 9$.

\subsubsection*{Note on the Computational Methods}
Most of the computations for this work were performed using the computer algebra system GAP \cite{GAP4}. The calculation of lattice automorphism groups and isomorphisms between lattices was a crucial part for which a modified version of the algorithm described in~\cite{Plesken:1997} was implemented. The computation of the Smith-Minkowski-Siegel mass formula relied on the built-in method \texttt{conway\char`_mass()} of the computer algebra system SAGE~\cite{Sage:2014}. Gr\"obner bases were calculated using Singular~\cite{Singular} and the GAP package ``singular".


\subsection*{Acknowledgements}
The author was supported by the Grant-in-Aid for Scientific Research from the Ministry of Education, Science, Sports, and Culture (MEXT), Japan (No. 23104011). He also wants to thank N. Maekawa, T. Kobayashi and S. Kuwakino for many valuable and inspiring discussions.


\printbibliography

\input{table_gram}

\begin{table}[p]
\centering
\input{table_bottom}
\caption{The bottom nodes of $\mathscr{G}$, i.e. the lattices $(\Gamma_{14})_\text{R}$ generated from the 32 elementary lattices $\Xi$. Here, $\Delta_2^\perp$ denotes the root system of norm 2 vectors orthogonal to $D_5^\text{st}$. Also, some information on the corresponding genera $\mathcal{G}_\text{L}$ is provided.}
\label{t:bottoms}
\end{table}

\begin{table}[p]
\centering
\input{table_top}
\caption{The top nodes in $\mathscr{G}$. The corresponding left-mover genera $\mathcal{G}_\text{L}$ were enumerated completely. Here, $\Gamma^\text{st}$ denotes the spacetime sublattice, and $\Delta_2^\perp$ is the root system of norm 2 vectors orthogonal to $\Gamma^\text{st}$.}
\label{t:tops}
\end{table}

\begin{table}[p]
\centering
\input{table_susy}
\caption{Right-mover lattices $(\Gamma_{14})_\text{R}$ with $\mathcal{N}=1$. Here, $\Delta_2^\perp$ denotes the root system of norm 2 vectors orthogonal to $E_6^\text{st}$. Also, some information on the corresponding genera $\mathcal{G}_\text{L}$ is provided. Where applicable, the corresponding $Z_N$ twist-orbifolds are shown.}
\label{t:susy}
\end{table}
 
\begin{sidewaysfigure}[p]
\centering
\adjustbox{scale = 0.87}{\input{C1.tex}}
\caption{Lattices $(\Gamma_{14})_\text{R}$ in $\mathscr{G}_1$ with spacetime supersymmetry. The node label indicates the root system of norm 2 vectors and the elementary divisors of $(\Gamma_{14})_\text{R}$.}
\label{f:C1}
\end{sidewaysfigure}
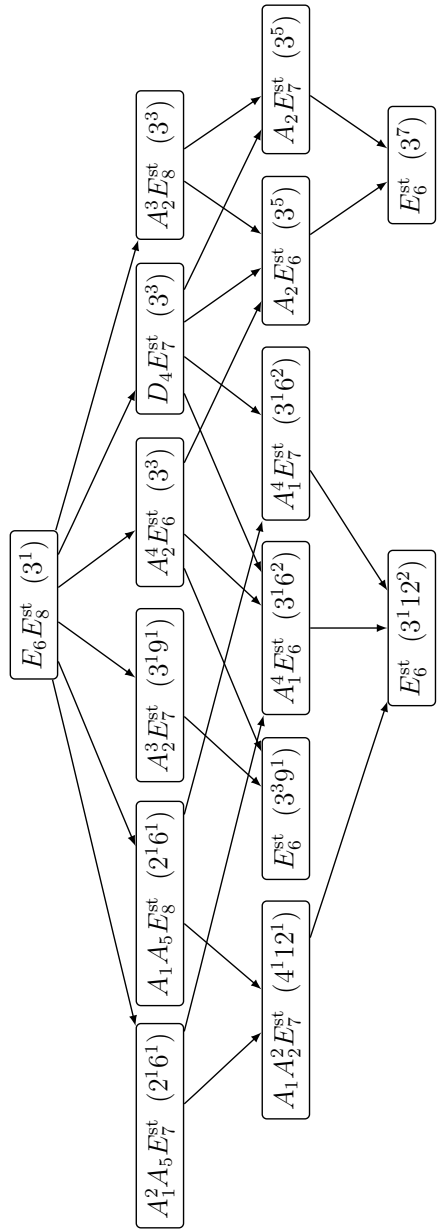

\begin{sidewaysfigure}[p]
\centering
\adjustbox{scale = 0.87}{\input{C4.tex}}
\caption{Lattices $(\Gamma_{14})_\text{R}$ in $\mathscr{G}_2$ with spacetime supersymmetry. The node label indicates the root system of norm 2 vectors and the elementary divisors of $(\Gamma_{14})_\text{R}$.}
\label{f:C2}
\end{sidewaysfigure}

\end{document}

%% file: table_gram.tex

\begin{table}[p]
\centering
\begin{tabular}{lclll}
\toprule
$\Xi$ & Gram Matrix & Divisors & $\Delta_2$ & $n_3$ \\
\midrule
$1$ & \scriptsize $\begin{pmatrix}3\end{pmatrix}$ & $3^{1}$ & - & $2$ \\
$3$ & \scriptsize $\begin{pmatrix}3 & -1 & -1\\
-1 & 3 & -1\\
-1 & -1 & 3\end{pmatrix}$ & $4^{2}$ & - & $8$ \\
$5$ & \scriptsize $\begin{pmatrix}3 & -1 & -1 & -1 & -1\\
-1 & 3 & -1 & -1 & 1\\
-1 & -1 & 3 & 1 & -1\\
-1 & -1 & 1 & 3 & -1\\
-1 & 1 & -1 & -1 & 3\end{pmatrix}$ & $2^{3}6^{1}$ & - & $20$ \\
$6_A$ & \scriptsize $\begin{pmatrix}3 & -1 & -1 & -1 & -1 & -1\\
-1 & 3 & -1 & -1 & -1 & 1\\
-1 & -1 & 3 & 1 & 1 & -1\\
-1 & -1 & 1 & 3 & 1 & -1\\
-1 & -1 & 1 & 1 & 3 & -1\\
-1 & 1 & -1 & -1 & -1 & 3\end{pmatrix}$ & $2^{4}4^{1}$ & - & $32$ \\
$6_B$ & \scriptsize $\begin{pmatrix}3 & -1 & -1 & -1 & -1 & -1\\
-1 & 3 & -1 & -1 & 1 & 1\\
-1 & -1 & 3 & 1 & -1 & 1\\
-1 & -1 & 1 & 3 & 0 & 0\\
-1 & 1 & -1 & 0 & 3 & 0\\
-1 & 1 & 1 & 0 & 0 & 3\end{pmatrix}$ & $2^{1}8^{2}$ & - & $20$ \\
$6_C$ & \scriptsize $\begin{pmatrix}3 & -1 & -1 & -1 & -1 & 0\\
-1 & 3 & -1 & 0 & 0 & -1\\
-1 & -1 & 3 & 0 & 1 & 0\\
-1 & 0 & 0 & 3 & -1 & -1\\
-1 & 0 & 1 & -1 & 3 & 0\\
0 & -1 & 0 & -1 & 0 & 3\end{pmatrix}$ & $5^{3}$  & - & $20$ \\
$7_A$ & \scriptsize $\begin{pmatrix}3 & -1 & -1 & -1 & -1 & -1 & -1\\
-1 & 3 & -1 & -1 & -1 & -1 & 1\\
-1 & -1 & 3 & 1 & 1 & 1 & -1\\
-1 & -1 & 1 & 3 & 1 & 1 & -1\\
-1 & -1 & 1 & 1 & 3 & 1 & -1\\
-1 & -1 & 1 & 1 & 1 & 3 & -1\\
-1 & 1 & -1 & -1 & -1 & -1 & 3\end{pmatrix}$ & $2^{6}$  & - & $56$ \\
$7_B$ & \scriptsize $\begin{pmatrix}3 & -1 & -1 & -1 & -1 & -1 & -1\\
-1 & 3 & -1 & -1 & -1 & 1 & 1\\
-1 & -1 & 3 & 1 & 1 & -1 & -1\\
-1 & -1 & 1 & 3 & 1 & -1 & 0\\
-1 & -1 & 1 & 1 & 3 & 0 & -1\\
-1 & 1 & -1 & -1 & 0 & 3 & 1\\
-1 & 1 & -1 & 0 & -1 & 1 & 3\end{pmatrix}$ & $12^{2}$ & - & $32$ \\
$8_A$ & \scriptsize $\begin{pmatrix}3 & -2 & -2 & -1 & -1 & -1 & -1 & -1\\
-2 & 3 & 1 & 0 & 0 & 0 & 0 & 1\\
-2 & 1 & 3 & 0 & 0 & 0 & 0 & 1\\
-1 & 0 & 0 & 3 & 1 & 1 & 1 & -1\\
-1 & 0 & 0 & 1 & 3 & 1 & 1 & -1\\
-1 & 0 & 0 & 1 & 1 & 3 & 1 & -1\\
-1 & 0 & 0 & 1 & 1 & 1 & 3 & -1\\
-1 & 1 & 1 & -1 & -1 & -1 & -1 & 3\end{pmatrix}$ & $2^{4}4^{1}$ & $A_1^2$ & $80$ \\
\bottomrule
\end{tabular}
\caption{Gram matrices for the primitive elementary lattices $\Xi$. Also shown are the elementary divisors, the root system $\Delta_2$ of norm 2 vectors and the number $n_3$ of norm 3 vectors.}
\label{tableGram}
\end{table}

\begin{table}[p]
\centering
\ContinuedFloat
\begin{tabular}{lclll}
\toprule
$\Xi$ & Gram Matrix & Divisors & $\Delta_2$ & $n_3$  \\
\midrule
$8_B$ & \scriptsize $\begin{pmatrix}3 & -1 & -1 & -1 & -1 & -1 & -1 & -1\\
-1 & 3 & -1 & -1 & -1 & -1 & 1 & 1\\
-1 & -1 & 3 & 1 & 1 & 1 & -1 & -1\\
-1 & -1 & 1 & 3 & 1 & 1 & -1 & 0\\
-1 & -1 & 1 & 1 & 3 & 1 & -1 & 0\\
-1 & -1 & 1 & 1 & 1 & 3 & 0 & -1\\
-1 & 1 & -1 & -1 & -1 & 0 & 3 & 1\\
-1 & 1 & -1 & 0 & 0 & -1 & 1 & 3\end{pmatrix}$ & $4^{2}12^{1}$ & - & $44$ \\
$8_C$ & \scriptsize $\begin{pmatrix}3 & -1 & -1 & -1 & -1 & -1 & -1 & -1\\
-1 & 3 & -1 & -1 & -1 & 1 & 1 & 1\\
-1 & -1 & 3 & 1 & 1 & -1 & -1 & 1\\
-1 & -1 & 1 & 3 & 1 & -1 & 0 & 0\\
-1 & -1 & 1 & 1 & 3 & 0 & 0 & 0\\
-1 & 1 & -1 & -1 & 0 & 3 & 1 & 0\\
-1 & 1 & -1 & 0 & 0 & 1 & 3 & 0\\
-1 & 1 & 1 & 0 & 0 & 0 & 0 & 3\end{pmatrix}$ & $2^{1}8^{1}24^{1}$ & - & $32$ \\
$8_D$ & \scriptsize $\begin{pmatrix}3 & -1 & -1 & -1 & -1 & -1 & -1 & 0\\
-1 & 3 & -1 & 0 & 0 & 0 & 0 & -1\\
-1 & -1 & 3 & 0 & 0 & 1 & 1 & 0\\
-1 & 0 & 0 & 3 & 1 & -1 & 0 & -1\\
-1 & 0 & 0 & 1 & 3 & 0 & -1 & 0\\
-1 & 0 & 1 & -1 & 0 & 3 & 1 & 1\\
-1 & 0 & 1 & 0 & -1 & 1 & 3 & -1\\
0 & -1 & 0 & -1 & 0 & 1 & -1 & 3\end{pmatrix}$ & $3^{2}6^{2}$ & - & $32$ \\
$9_A$ & \scriptsize $\begin{pmatrix}3 & -2 & -2 & -1 & -1 & -1 & -1 & -1 & -1\\
-2 & 3 & 1 & 0 & 0 & 0 & 0 & 0 & 1\\
-2 & 1 & 3 & 0 & 0 & 0 & 0 & 0 & 1\\
-1 & 0 & 0 & 3 & 1 & 1 & 1 & 1 & -1\\
-1 & 0 & 0 & 1 & 3 & 1 & 1 & 1 & -1\\
-1 & 0 & 0 & 1 & 1 & 3 & 1 & 1 & -1\\
-1 & 0 & 0 & 1 & 1 & 1 & 3 & 1 & -1\\
-1 & 0 & 0 & 1 & 1 & 1 & 1 & 3 & 0\\
-1 & 1 & 1 & -1 & -1 & -1 & -1 & 0 & 3\end{pmatrix}$ & $2^{4}8^{1}$ & $A_1^2$ & $88$ \\
$9_B$ & \scriptsize $\begin{pmatrix}3 & -2 & -2 & -1 & -1 & -1 & -1 & -1 & -1\\
-2 & 3 & 1 & 0 & 0 & 0 & 0 & 1 & 1\\
-2 & 1 & 3 & 0 & 0 & 0 & 0 & 1 & 1\\
-1 & 0 & 0 & 3 & 1 & 1 & 1 & -1 & -1\\
-1 & 0 & 0 & 1 & 3 & 1 & 1 & -1 & 0\\
-1 & 0 & 0 & 1 & 1 & 3 & 1 & -1 & 0\\
-1 & 0 & 0 & 1 & 1 & 1 & 3 & -1 & 0\\
-1 & 1 & 1 & -1 & -1 & -1 & -1 & 3 & 1\\
-1 & 1 & 1 & -1 & 0 & 0 & 0 & 1 & 3\end{pmatrix}$ & $2^{2}4^{1}8^{1}$ & $A_1^2$ & $84$\\
$9_C$ & \scriptsize $\begin{pmatrix}3 & -2 & -2 & -1 & -1 & -1 & -1 & -1 & 0\\
-2 & 3 & 1 & 0 & 0 & 0 & 0 & 1 & 0\\
-2 & 1 & 3 & 0 & 0 & 0 & 0 & 1 & 0\\
-1 & 0 & 0 & 3 & 1 & 1 & 1 & -1 & 0\\
-1 & 0 & 0 & 1 & 3 & 1 & 1 & -1 & 0\\
-1 & 0 & 0 & 1 & 1 & 3 & 1 & -1 & 0\\
-1 & 0 & 0 & 1 & 1 & 1 & 3 & 0 & 0\\
-1 & 1 & 1 & -1 & -1 & -1 & 0 & 3 & -1\\
0 & 0 & 0 & 0 & 0 & 0 & 0 & -1 & 3\end{pmatrix}$ & $2^{2}8^{2}$ & $A_1^2$ & $60$\\
\bottomrule
\end{tabular}
\caption{The lattices $\Xi$ (cont'd).}
\label{t:gram}
\end{table}

\begin{table}[p]
\centering
\ContinuedFloat
\begin{tabular}{lclll}
\toprule
$\Xi$ & Gram Matrix & Divisors & $\Delta_2$ & $n_3$ \\
\midrule
$9_D$ & \scriptsize $\begin{pmatrix}3 & -2 & -1 & -1 & -1 & -1 & -1 & -1 & -1\\
-2 & 3 & 0 & 0 & 0 & 0 & 0 & 0 & 1\\
-1 & 0 & 3 & 0 & 0 & 0 & 0 & 0 & 0\\
-1 & 0 & 0 & 3 & 0 & 0 & 0 & 0 & 1\\
-1 & 0 & 0 & 0 & 3 & 0 & 1 & 1 & -1\\
-1 & 0 & 0 & 0 & 0 & 3 & 1 & 1 & 0\\
-1 & 0 & 0 & 0 & 1 & 1 & 3 & 0 & -1\\
-1 & 0 & 0 & 0 & 1 & 1 & 0 & 3 & 0\\
-1 & 1 & 0 & 1 & -1 & 0 & -1 & 0 & 3\end{pmatrix}$ & $3^{1}9^{2}$ & $A_1$ & $54$\\
$9_E$ & \scriptsize $\begin{pmatrix}3 & -1 & -1 & -1 & -1 & -1 & -1 & -1 & -1\\
-1 & 3 & -1 & -1 & -1 & -1 & -1 & 1 & 1\\
-1 & -1 & 3 & 1 & 1 & 1 & 1 & -1 & -1\\
-1 & -1 & 1 & 3 & 1 & 1 & 1 & -1 & 0\\
-1 & -1 & 1 & 1 & 3 & 1 & 1 & -1 & 0\\
-1 & -1 & 1 & 1 & 1 & 3 & 1 & -1 & 0\\
-1 & -1 & 1 & 1 & 1 & 1 & 3 & 0 & 0\\
-1 & 1 & -1 & -1 & -1 & -1 & 0 & 3 & 1\\
-1 & 1 & -1 & 0 & 0 & 0 & 0 & 1 & 3\end{pmatrix}$ & $2^{2}8^{2}$ & - & $64$\\
$9_F$ & \scriptsize $\begin{pmatrix}3 & -1 & -1 & -1 & -1 & -1 & -1 & -1 & -1\\
-1 & 3 & -1 & -1 & -1 & -1 & -1 & 1 & 1\\
-1 & -1 & 3 & 1 & 1 & 1 & 1 & -1 & -1\\
-1 & -1 & 1 & 3 & 1 & 1 & 1 & -1 & 0\\
-1 & -1 & 1 & 1 & 3 & 1 & 1 & -1 & 0\\
-1 & -1 & 1 & 1 & 1 & 3 & 1 & 0 & -1\\
-1 & -1 & 1 & 1 & 1 & 1 & 3 & 0 & -1\\
-1 & 1 & -1 & -1 & -1 & 0 & 0 & 3 & 1\\
-1 & 1 & -1 & 0 & 0 & -1 & -1 & 1 & 3\end{pmatrix}$ & $4^{4}$& - & $56$\\
$9_G$ & \scriptsize $\begin{pmatrix}3 & -1 & -1 & -1 & -1 & -1 & -1 & -1 & 0\\
-1 & 3 & -1 & -1 & -1 & -1 & 1 & 1 & 0\\
-1 & -1 & 3 & 1 & 1 & 1 & -1 & -1 & 0\\
-1 & -1 & 1 & 3 & 1 & 1 & -1 & 0 & -1\\
-1 & -1 & 1 & 1 & 3 & 1 & -1 & 0 & -1\\
-1 & -1 & 1 & 1 & 1 & 3 & 0 & -1 & -1\\
-1 & 1 & -1 & -1 & -1 & 0 & 3 & 1 & 1\\
-1 & 1 & -1 & 0 & 0 & -1 & 1 & 3 & 1\\
0 & 0 & 0 & -1 & -1 & -1 & 1 & 1 & 3\end{pmatrix}$ & $4^{2}16^{1}$& - & $56$\\
$9_H$ & \scriptsize $\begin{pmatrix}3 & -1 & -1 & -1 & -1 & -1 & -1 & -1 & -1\\
-1 & 3 & -1 & -1 & -1 & -1 & 1 & 1 & 1\\
-1 & -1 & 3 & 1 & 1 & 1 & -1 & -1 & 1\\
-1 & -1 & 1 & 3 & 1 & 1 & -1 & 0 & 0\\
-1 & -1 & 1 & 1 & 3 & 1 & -1 & 0 & 0\\
-1 & -1 & 1 & 1 & 1 & 3 & 0 & 0 & 0\\
-1 & 1 & -1 & -1 & -1 & 0 & 3 & 1 & 0\\
-1 & 1 & -1 & 0 & 0 & 0 & 1 & 3 & 0\\
-1 & 1 & 1 & 0 & 0 & 0 & 0 & 0 & 3\end{pmatrix}$ & $2^{1}4^{1}8^{2}$& - & $44$\\
$9_I$ & \scriptsize $\begin{pmatrix}3 & -1 & -1 & -1 & -1 & -1 & -1 & -1 & 0\\
-1 & 3 & -1 & -1 & -1 & -1 & 1 & 1 & 0\\
-1 & -1 & 3 & 1 & 1 & 1 & -1 & -1 & 0\\
-1 & -1 & 1 & 3 & 1 & 1 & -1 & 0 & -1\\
-1 & -1 & 1 & 1 & 3 & 1 & -1 & 0 & -1\\
-1 & -1 & 1 & 1 & 1 & 3 & 0 & 0 & 0\\
-1 & 1 & -1 & -1 & -1 & 0 & 3 & 1 & 1\\
-1 & 1 & -1 & 0 & 0 & 0 & 1 & 3 & 0\\
0 & 0 & 0 & -1 & -1 & 0 & 1 & 0 & 3\end{pmatrix}$ & $8^{3}$& - & $44$\\
\bottomrule
\end{tabular}
\caption{The lattices $\Xi$ (cont'd).}
\end{table}

\begin{table}[p]
\centering
\ContinuedFloat
\begin{tabular}{lclll}
\toprule
$\Xi$ & Gram Matrix & Divisors & $\Delta_2$ & $n_3$ \\
\midrule
$9_J$ & \scriptsize $\begin{pmatrix}3 & -1 & -1 & -1 & -1 & -1 & -1 & -1 & -1\\
-1 & 3 & -1 & -1 & -1 & -1 & 1 & 1 & 1\\
-1 & -1 & 3 & 1 & 1 & 1 & -1 & -1 & 0\\
-1 & -1 & 1 & 3 & 1 & 1 & -1 & 0 & -1\\
-1 & -1 & 1 & 1 & 3 & 1 & 0 & -1 & -1\\
-1 & -1 & 1 & 1 & 1 & 3 & 0 & 0 & 0\\
-1 & 1 & -1 & -1 & 0 & 0 & 3 & 1 & 1\\
-1 & 1 & -1 & 0 & -1 & 0 & 1 & 3 & 1\\
-1 & 1 & 0 & -1 & -1 & 0 & 1 & 1 & 3\end{pmatrix}$ & $3^{1}12^{2}$ & - & $44$\\
$9_K$ & \scriptsize $\begin{pmatrix}3 & -1 & -1 & -1 & -1 & -1 & -1 & -1 & -1\\
-1 & 3 & -1 & -1 & -1 & 0 & 0 & 0 & 0\\
-1 & -1 & 3 & 1 & 1 & 0 & 0 & 0 & 1\\
-1 & -1 & 1 & 3 & 1 & 0 & 0 & 1 & 0\\
-1 & -1 & 1 & 1 & 3 & 0 & 1 & 0 & 0\\
-1 & 0 & 0 & 0 & 0 & 3 & -1 & -1 & -1\\
-1 & 0 & 0 & 0 & 1 & -1 & 3 & 1 & 1\\
-1 & 0 & 0 & 1 & 0 & -1 & 1 & 3 & 1\\
-1 & 0 & 1 & 0 & 0 & -1 & 1 & 1 & 3\end{pmatrix}$ & $3^{3}9^{1}$ & - & $56$\\
$9_L$ & \scriptsize $\begin{pmatrix}3 & -1 & -1 & -1 & -1 & -1 & -1 & 0 & 0\\
-1 & 3 & -1 & -1 & -1 & 1 & 1 & 0 & 0\\
-1 & -1 & 3 & 1 & 1 & -1 & -1 & 0 & 0\\
-1 & -1 & 1 & 3 & 1 & -1 & 0 & -1 & -1\\
-1 & -1 & 1 & 1 & 3 & 0 & -1 & -1 & 1\\
-1 & 1 & -1 & -1 & 0 & 3 & 1 & 0 & 0\\
-1 & 1 & -1 & 0 & -1 & 1 & 3 & 0 & 0\\
0 & 0 & 0 & -1 & -1 & 0 & 0 & 3 & 0\\
0 & 0 & 0 & -1 & 1 & 0 & 0 & 0 & 3\end{pmatrix}$ & $20^{2}$ & - & $44$\\
$9_M$ & \scriptsize $\begin{pmatrix}3 & -1 & -1 & -1 & -1 & -1 & -1 & 0 & 0\\
-1 & 3 & -1 & -1 & -1 & 1 & 1 & 0 & 0\\
-1 & -1 & 3 & 1 & 1 & -1 & 1 & 0 & 0\\
-1 & -1 & 1 & 3 & 1 & 0 & 0 & -1 & -1\\
-1 & -1 & 1 & 1 & 3 & 0 & 0 & -1 & 1\\
-1 & 1 & -1 & 0 & 0 & 3 & 0 & 0 & 0\\
-1 & 1 & 1 & 0 & 0 & 0 & 3 & 0 & 0\\
0 & 0 & 0 & -1 & -1 & 0 & 0 & 3 & 0\\
0 & 0 & 0 & -1 & 1 & 0 & 0 & 0 & 3\end{pmatrix}$ & $4^{2}8^{2}$ & - & $32$\\
$9_N$ & \scriptsize $\begin{pmatrix}3 & -1 & -1 & -1 & -1 & -1 & -1 & -1 & -1\\
-1 & 3 & -1 & -1 & 0 & 0 & 0 & 0 & 0\\
-1 & -1 & 3 & 1 & 0 & 0 & 0 & 1 & 1\\
-1 & -1 & 1 & 3 & 0 & 0 & 1 & 0 & 1\\
-1 & 0 & 0 & 0 & 3 & 0 & -1 & -1 & 1\\
-1 & 0 & 0 & 0 & 0 & 3 & 0 & 0 & 1\\
-1 & 0 & 0 & 1 & -1 & 0 & 3 & 1 & -1\\
-1 & 0 & 1 & 0 & -1 & 0 & 1 & 3 & 0\\
-1 & 0 & 1 & 1 & 1 & 1 & -1 & 0 & 3\end{pmatrix}$ & $7^{3}$ & - & $44$\\
\bottomrule
\end{tabular}
\caption{The lattices $\Xi$ (cont'd).}
\end{table}

%% file: table_bottom.tex

\begin{tabular}{cllll}
\toprule
  & \multicolumn{2}{c}{Genus} & \multicolumn{2}{c}{Lattice $(\Gamma_{14})_\text{R}$}  \\
\cmidrule(rl){2-3}\cmidrule(rl){4-5}
Component  &  $|\mathcal{G}_\text{L}|$ Bound  & Divisors & $\Xi$ & $\Delta_2^\perp$\\
\midrule
\multirow{6}{*}{$\mathscr{G}_1$} & $3.1\cdot 10^{17}$ & $3^7 6^2$ & $1^9$ & -   \\ 
								& $1.3\cdot 10^{9}$ & $2^4 4^1 12^1$ & $1^1 8_A^1$ & $A_1^2$   \\
								& $4.2\cdot 10^{14}$ & $3^1 6^4$ & $1^1 8_D^1$ & -   \\
								& $3.4\cdot 10^{12}$ & $3^1 18^2$ & $9_D^1$ & $A_1$   \\
								& $1.0\cdot 10^{14}$ & $2^1 6^1 12^2$ & $9_J^1$ & -   \\
								& $2.9\cdot 10^{11}$ & $3^2 6^1 18^1$ & $9_K^1$ & -   \\
\midrule
\multirow{11}{*}{$\mathscr{G}_2$}& $5.8\cdot 10^{17}$ & $6^2 12^2$ & $1^2 7_B^1$ & -   \\
								& $3.8\cdot 10^{17}$ & $2^6 12^2$ & $1^1 3^1 5^1$ & -   \\		
								& $2.1\cdot 10^{15}$ & $4^2 12^2$ & $1^1 8_B^1$ & -   \\
								& $1.8\cdot 10^{15}$ & $2^2 4^6$ & $3^3$ & -   \\
								& $1.2\cdot 10^{12}$ & $2^4 4^4$ & $3^1 6_A^1$ & -   \\
								& $2.1\cdot 10^{11}$ & $2^4 8^2$ & $9_C^1$ & $A_1^2$   \\
								& $5.2\cdot 10^{9}$ & $2^4 8^2$ & $9_E^1$ & -   \\
								& $4.4\cdot 10^{9}$ & $2^2 4^4$ & $9_F^1$ & -   \\
								& $4.2\cdot 10^{11}$ & $4^3 16^1$ & $9_G^1$ & -   \\
								& $3.6\cdot 10^{14}$ & $2^2 20^2$ & $9_L^1$ & -   \\
								&  $1.7\cdot 10^{15}$ & $2^2 4^2 8^2$ & $9_M^1$ & -   \\
\midrule
\multirow{1}{*}{$\mathscr{G}_3$} & $4.1\cdot 10^{13}$ & $7^1 14^2$ & $9_N^1$  & -   \\
\midrule
\multirow{6}{*}{$\mathscr{G}_4$} & $6.1\cdot 10^{18}$ & $2^1 4^1 24^2$ & $1^1 8_C^1$& -   \\
								& $8.4\cdot 10^{17}$ & $2^1 4^3 8^2$ & $3^1 6_B^1$ & -   \\
								
								& \multirow{2}{*}{$1.9\cdot 10^{7}$} &  \multirow{2}{*}{$2^4 4^1 8^1$} & $9_A^1$ & $A_1^2$   \\
								&									 & & $9_B^1$ & $A_1^2$   \\
								& $5.1\cdot 10^{13}$ & $2^3 4^1 8^2$  & $9_H^1$ & -    \\
								& $3.0\cdot 10^{14}$ & $4^1 8^3$ & $9_I^1$ & -    \\
\midrule
\multirow{3}{*}{$\mathscr{G}_5$} & $1.2\cdot 10^{18}$ & $2^1 6^5$ & $1^4 5^1$ & -   \\
								& $8.2\cdot 10^{21}$ & $2^2 4^1 12^3$ & $1^3 3^2$ & -   \\
								& $6.0\cdot 10^{17}$ & $2^3 6^1 12^2$ & $1^3 6_A^1$ & -   \\
\midrule
\multirow{1}{*}{$\mathscr{G}_6$} & $6.5\cdot 10^{22}$ & $15^1 30^2$ & $1^3 6_C^1$ & -   \\
\midrule

\multirow{1}{*}{$\mathscr{G}_7$} & $2.6\cdot 10^{20}$ & $2^1 10^1 20^2$ & $3^1 6_C^1$ & -  \\
\midrule
\multirow{1}{*}{$\mathscr{G}_8$}& $5.2\cdot 10^{22}$ & $2^1 12^1 24^2$ & $1^3 6_B^1$ & -   \\
\midrule
\multirow{2}{*}{$\mathscr{G}_9$} & $2.7\cdot 10^{23}$ & $3^2 6^2 12^2$ & $1^6 3^1$ & -   \\
								& $5.6\cdot 10^{11}$ & $2^6 6^2$ & $1^2 7_A^1$ & -   \\
\bottomrule
\end{tabular}

%% file: table_top.tex

\begin{tabular}{crlll}
\toprule
 & \multicolumn{2}{c}{Genus} & \multicolumn{2}{c}{Lattice $(\Gamma_{14})_\text{R}$} \\
\cmidrule(rl){2-3}\cmidrule(rl){4-5}
Component & $|\mathcal{G}_\text{L}|$ & Divisors &  $\Delta_2^\perp$ & $\Gamma^\text{st}$  \\
\midrule 
\multirow{2}{*}{$\mathscr{G}_1$} & \multirow{2}{*}{$31$} & \multirow{2}{*}{$3^1$} & $E_6$ & $E_8^\text{st}$  \\
								& & &  - & $D_{13}^\text{st}$ \\
\midrule
\multirow{3}{*}{$\mathscr{G}_2$} & \multirow{3}{*}{$68$}  & \multirow{3}{*}{$2^2$} & $D_6$ & $E_8^\text{st}$ \\
								& & & $A_1^2$ & $D_{12}^\text{st}$ \\					
								& & & - & $D_{14}^\text{st}$ \\
\midrule
\multirow{1}{*}{$\mathscr{G}_3$} & \multirow{1}{*}{$153$} & \multirow{1}{*}{$7^1$} & $A_6$ & $E_8^\text{st}$ \\
\midrule
\multirow{3}{*}{$\mathscr{G}_4$} &\multirow{3}{*}{$326$} & \multirow{3}{*}{$2^1 4^1$} & $A_1D_5$ & $E_8^\text{st}$ \\
								& & & $A_1A_3$ & $D_{10}^\text{st}$ \\
								& & & $A_1$ & $D_{13}^\text{st}$ \\
\midrule
\multirow{3}{*}{$\mathscr{G}_5$} &  \multirow{3}{*}{$382$} &  \multirow{3}{*}{$2^1 6^1$} &  $A_2D_4$ & $E_8^\text{st}$ \\
								& & & $A_1^3$ & $D_{10}^\text{st}$  \\
								& & & $A_2$ & $D_{12}^\text{st}$ \\
\midrule
\multirow{1}{*}{$\mathscr{G}_6$} & \multirow{1}{*}{$1163$}& \multirow{1}{*}{$15^1$} & $A_2A_4$ & $E_8^\text{st}$ \\
\midrule
\multirow{2}{*}{$\mathscr{G}_7$} & \multirow{2}{*}{$4043$} & \multirow{2}{*}{$2^1 10^1$} & $A_1^2 A_4$ & $E_8^\text{st}$ \\
								& & & $A_4$ & $D_{10}^\text{st}$ \\
\midrule
\multirow{2}{*}{$\mathscr{G}_8$} &\multirow{2}{*}{$9346$} & \multirow{2}{*}{$2^1 12^1$} &  $A_1A_2A_3$  & $E_8^\text{st}$ \\
								& & & $A_1A_2$ & $D_{11}^\text{st}$\\
\midrule
\multirow{4}{*}{$\mathscr{G}_9$} & \multirow{4}{*}{$19832$} & \multirow{4}{*}{$6^2$} & $A_1^2A_2^2$ & $E_8^\text{st}$\\
								& & &   $A_1^2$ & $D_{10}^\text{st}$ \\
								& & &  $A_2^2$ & $D_{10}^\text{st}$  \\
								& & &  - & $D_{12}^\text{st}$  \\
\bottomrule
\end{tabular}

%% file: table_susy.tex

\begin{tabular}{cllll}
\toprule
 & \multicolumn{2}{c}{Genus} & \multicolumn{2}{c}{Lattice $(\Gamma_{14})_\text{R}$}\\
\cmidrule(rl){2-3}\cmidrule(rl){4-5}
Component & Divisors &  $|\mathcal{G}_\text{L}|$ & $\Delta_2^\perp$ & $(\Gamma_{6})_\text{R}/Z_N$ Orbifolds \\
 \midrule
 \multirow{6}{*}{$\mathscr{G}_1$}  & $3^3$ & 2030 & $A_2^4$  & $E_6/{Z}_3$ \\
  & $3^1 6^2$ & $>1.5 \cdot 10^3$ & $A_1^4$ & $E_6/{Z}_6^\text{I}, E_6/{Z}_6^\text{II}$ \\
 & $3^5$ & $>6.9 \cdot 10^3$ & $A_2$ & $A_2^3/{Z}_3$\\
 & $3^3 9^1$ & $>2.7 \cdot 10^5$ & - &  \\
 & $3^1 12^2$ & $>1.5\cdot10^9$ & - & $E_6/{Z}_{12}^{\text{I}}, A_1A_5/{Z}_6^\text{II}$\\
 & $3^7$ & $>4.1 \cdot 10^8$ & - &  \\
  \midrule
 \multirow{9}{*}{$\mathscr{G}_2$}  & $2^6$ & $-$ & $A_3$ & \\
 & $2^24^2$ & $>3$ & $A_1^4$ & \\
 & $2^24^2$ & $>6$ & $A_1^6$ & $D_6/{Z}_4$\\
 & $2^44^2$ & $>1.3 \cdot 10^5$ & $A_1^2$ & $A_1^2D_4/{Z}_4$ \\
 & $2^44^2$ & $>1.3 \cdot 10^4$ & - & \\
 & $2^28^2$ & $>4.8 \cdot 10^6$ & $A_1$ & $A_3^2/{Z}_4, D_6/{Z}_8^\text{I}$ \\
 & $2^28^2$ & $>8.0 \cdot 10^5$ & - & \\
 & $4^4$ & $>8.0 \cdot 10^5$ & - & \\
 & $2^24^4$ & $>1.7 \cdot 10^9$ & - & \\
  \midrule
  \multirow{1}{*}{$\mathscr{G}_3$} & $7^3$ & $> 4.0 \cdot 10^7$ & -& $A_6/Z_7$ \\
 \midrule
  \multirow{2}{*}{$\mathscr{G}_4$} & $2^24^18^1$ & $> 4.4 \cdot 10^3$ & $A_1^2$ &  \\
  & $2^14^18^2$ & $> 2.3 \cdot 10^9$ & - & $A_1D_5/{Z}_8^\text{II}$ \\
   \midrule
 \multirow{1}{*}{$\mathscr{G}_5$} & $2^16^3$ & $>5.2 \cdot 10^7$ & - & $A_2D_4/{Z}_6^\text{II}$ \\
\bottomrule 
\end{tabular}

%% file: C1.tex

\begin{tikzpicture}[>=latex,line join=bevel,]
\begin{scope}
  \pgfsetstrokecolor{black}
  \definecolor{strokecol}{rgb}{1.0,1.0,1.0};
  \pgfsetstrokecolor{strokecol}
  \definecolor{fillcol}{rgb}{1.0,1.0,1.0};
  \pgfsetfillcolor{fillcol}
  \filldraw (0bp,0bp) -- (0bp,180bp) -- (519bp,180bp) -- (519bp,0bp) -- cycle;
\end{scope}
\begin{scope}
  \pgfsetstrokecolor{black}
  \definecolor{strokecol}{rgb}{1.0,1.0,1.0};
  \pgfsetstrokecolor{strokecol}
  \definecolor{fillcol}{rgb}{1.0,1.0,1.0};
  \pgfsetfillcolor{fillcol}
  \filldraw (0bp,0bp) -- (0bp,180bp) -- (519bp,180bp) -- (519bp,0bp) -- cycle;
\end{scope}
\begin{scope}
  \pgfsetstrokecolor{black}
  \definecolor{strokecol}{rgb}{1.0,1.0,1.0};
  \pgfsetstrokecolor{strokecol}
  \definecolor{fillcol}{rgb}{1.0,1.0,1.0};
  \pgfsetfillcolor{fillcol}
  \filldraw (0bp,0bp) -- (0bp,180bp) -- (519bp,180bp) -- (519bp,0bp) -- cycle;
\end{scope}
\begin{scope}
  \pgfsetstrokecolor{black}
  \definecolor{strokecol}{rgb}{1.0,1.0,1.0};
  \pgfsetstrokecolor{strokecol}
  \definecolor{fillcol}{rgb}{1.0,1.0,1.0};
  \pgfsetfillcolor{fillcol}
  \filldraw (0bp,0bp) -- (0bp,180bp) -- (519bp,180bp) -- (519bp,0bp) -- cycle;
\end{scope}
\begin{scope}
  \pgfsetstrokecolor{black}
  \definecolor{strokecol}{rgb}{1.0,1.0,1.0};
  \pgfsetstrokecolor{strokecol}
  \definecolor{fillcol}{rgb}{1.0,1.0,1.0};
  \pgfsetfillcolor{fillcol}
  \filldraw (0bp,0bp) -- (0bp,180bp) -- (519bp,180bp) -- (519bp,0bp) -- cycle;
\end{scope}
\begin{scope}
  \pgfsetstrokecolor{black}
  \definecolor{strokecol}{rgb}{1.0,1.0,1.0};
  \pgfsetstrokecolor{strokecol}
  \definecolor{fillcol}{rgb}{1.0,1.0,1.0};
  \pgfsetfillcolor{fillcol}
  \filldraw (0bp,0bp) -- (0bp,180bp) -- (519bp,180bp) -- (519bp,0bp) -- cycle;
\end{scope}
\begin{scope}
  \pgfsetstrokecolor{black}
  \definecolor{strokecol}{rgb}{1.0,1.0,1.0};
  \pgfsetstrokecolor{strokecol}
  \definecolor{fillcol}{rgb}{1.0,1.0,1.0};
  \pgfsetfillcolor{fillcol}
  \filldraw (0bp,0bp) -- (0bp,180bp) -- (519bp,180bp) -- (519bp,0bp) -- cycle;
\end{scope}
\begin{scope}
  \pgfsetstrokecolor{black}
  \definecolor{strokecol}{rgb}{1.0,1.0,1.0};
  \pgfsetstrokecolor{strokecol}
  \definecolor{fillcol}{rgb}{1.0,1.0,1.0};
  \pgfsetfillcolor{fillcol}
  \filldraw (0bp,0bp) -- (0bp,180bp) -- (519bp,180bp) -- (519bp,0bp) -- cycle;
\end{scope}
\begin{scope}
  \pgfsetstrokecolor{black}
  \definecolor{strokecol}{rgb}{1.0,1.0,1.0};
  \pgfsetstrokecolor{strokecol}
  \definecolor{fillcol}{rgb}{1.0,1.0,1.0};
  \pgfsetfillcolor{fillcol}
  \filldraw (0bp,0bp) -- (0bp,180bp) -- (519bp,180bp) -- (519bp,0bp) -- cycle;
\end{scope}
\begin{scope}
  \pgfsetstrokecolor{black}
  \definecolor{strokecol}{rgb}{1.0,1.0,1.0};
  \pgfsetstrokecolor{strokecol}
  \definecolor{fillcol}{rgb}{1.0,1.0,1.0};
  \pgfsetfillcolor{fillcol}
  \filldraw (0bp,0bp) -- (0bp,180bp) -- (519bp,180bp) -- (519bp,0bp) -- cycle;
\end{scope}
\begin{scope}
  \pgfsetstrokecolor{black}
  \definecolor{strokecol}{rgb}{1.0,1.0,1.0};
  \pgfsetstrokecolor{strokecol}
  \definecolor{fillcol}{rgb}{1.0,1.0,1.0};
  \pgfsetfillcolor{fillcol}
  \filldraw (0bp,0bp) -- (0bp,180bp) -- (519bp,180bp) -- (519bp,0bp) -- cycle;
\end{scope}
\begin{scope}
  \pgfsetstrokecolor{black}
  \definecolor{strokecol}{rgb}{1.0,1.0,1.0};
  \pgfsetstrokecolor{strokecol}
  \definecolor{fillcol}{rgb}{1.0,1.0,1.0};
  \pgfsetfillcolor{fillcol}
  \filldraw (0bp,0bp) -- (0bp,180bp) -- (519bp,180bp) -- (519bp,0bp) -- cycle;
\end{scope}
\begin{scope}
  \pgfsetstrokecolor{black}
  \definecolor{strokecol}{rgb}{1.0,1.0,1.0};
  \pgfsetstrokecolor{strokecol}
  \definecolor{fillcol}{rgb}{1.0,1.0,1.0};
  \pgfsetfillcolor{fillcol}
  \filldraw (0bp,0bp) -- (0bp,180bp) -- (519bp,180bp) -- (519bp,0bp) -- cycle;
\end{scope}
\begin{scope}
  \pgfsetstrokecolor{black}
  \definecolor{strokecol}{rgb}{1.0,1.0,1.0};
  \pgfsetstrokecolor{strokecol}
  \definecolor{fillcol}{rgb}{1.0,1.0,1.0};
  \pgfsetfillcolor{fillcol}
  \filldraw (0bp,0bp) -- (0bp,180bp) -- (519bp,180bp) -- (519bp,0bp) -- cycle;
\end{scope}
\begin{scope}
  \pgfsetstrokecolor{black}
  \definecolor{strokecol}{rgb}{1.0,1.0,1.0};
  \pgfsetstrokecolor{strokecol}
  \definecolor{fillcol}{rgb}{1.0,1.0,1.0};
  \pgfsetfillcolor{fillcol}
  \filldraw (0bp,0bp) -- (0bp,180bp) -- (519bp,180bp) -- (519bp,0bp) -- cycle;
\end{scope}
\begin{scope}[rounded corners=2pt,semithick]
  \node (l14) at (135bp,117bp) [draw,rectangle] {$A_{1}A_{5}E_{8}^{\text{st}}\;\; (2^{1}6^{1})$};
  \node (l15) at (454bp,117bp) [draw,rectangle] {$A_{2}^{3}E_{8}^{\text{st}}\;\; (3^{3})$};
  \node (l10) at (417bp,63bp) [draw,rectangle] {$A_{2}E_{6}^{\text{st}}\;\; (3^{5})$};
  \node (l11) at (39bp,117bp) [draw,rectangle] {$A_{1}^{2}A_{5}E_{7}^{\text{st}}\;\; (2^{1}6^{1})$};
  \node (l12) at (379bp,117bp) [draw,rectangle] {$D_{4}E_{7}^{\text{st}}\;\; (3^{3})$};
  \node (l13) at (264bp,171bp) [draw,rectangle] {$E_{6}E_{8}^{\text{st}}\;\; (3^{1})$};
  \node (l6) at (225bp,117bp) [draw,rectangle] {$A_{2}^{3}E_{7}^{\text{st}}\;\; (3^{1}9^{1})$};
  \node (l7) at (338bp,63bp) [draw,rectangle] {$A_{1}^{4}E_{7}^{\text{st}}\;\; (3^{1}6^{2})$};
  \node (l4) at (254bp,63bp) [draw,rectangle] {$A_{1}^{4}E_{6}^{\text{st}}\;\; (3^{1}6^{2})$};
  \node (l5) at (304bp,117bp) [draw,rectangle] {$A_{2}^{4}E_{6}^{\text{st}}\;\; (3^{3})$};
  \node (l2) at (176bp,63bp) [draw,rectangle] {$E_{6}^{\text{st}}\;\; (3^{3}9^{1})$};
  \node (l3) at (454bp,9bp) [draw,rectangle] {$E_{6}^{\text{st}}\;\; (3^{7})$};
  \node (l1) at (254bp,9bp) [draw,rectangle] {$E_{6}^{\text{st}}\;\; (3^{1}12^{2})$};
  \node (l8) at (89bp,63bp) [draw,rectangle] {$A_{1}A_{2}^{2}E_{7}^{\text{st}}\;\; (4^{1}12^{1})$};
  \node (l9) at (491bp,63bp) [draw,rectangle] {$A_{2}E_{7}^{\text{st}}\;\; (3^{5})$};
\end{scope}
\begin{scope}[semithick]
  \draw [->] (l10) ..controls (427.96bp,46.591bp) and (435.73bp,35.679bp)  .. (l3);
  \draw [->] (l12) ..controls (366.79bp,100.52bp) and (358.07bp,89.452bp)  .. (l7);
  \draw [->] (l14) ..controls (121.23bp,100.44bp) and (111.31bp,89.223bp)  .. (l8);
  \draw [->] (l15) ..controls (464.96bp,100.59bp) and (472.73bp,89.679bp)  .. (l9);
  \draw [->] (l14) ..controls (201.29bp,99.019bp) and (258.18bp,84.445bp)  .. (l7);
  \draw [->] (l13) ..controls (275.91bp,154.52bp) and (284.42bp,143.45bp)  .. (l5);
  \draw [->] (l13) ..controls (300.7bp,153.4bp) and (330.29bp,140.02bp)  .. (l12);
  \draw [->] (l9) ..controls (480.04bp,46.591bp) and (472.27bp,35.679bp)  .. (l3);
  \draw [->] (l5) ..controls (288.96bp,100.36bp) and (278.03bp,88.994bp)  .. (l4);
  \draw [->] (l12) ..controls (414.66bp,99.442bp) and (443.3bp,86.145bp)  .. (l9);
  \draw [->] (l12) ..controls (339.25bp,99.464bp) and (306.59bp,85.877bp)  .. (l4);
  \draw [->] (l12) ..controls (390.26bp,100.59bp) and (398.23bp,89.679bp)  .. (l10);
  \draw [->] (l5) ..controls (340.07bp,99.403bp) and (369.14bp,86.025bp)  .. (l10);
  \draw [->] (l13) ..controls (324.55bp,153.43bp) and (379.73bp,138.33bp)  .. (l15);
  \draw [->] (l11) ..controls (109.37bp,98.981bp) and (169.98bp,84.321bp)  .. (l4);
  \draw [->] (l13) ..controls (198.34bp,154.82bp) and (129.24bp,138.86bp)  .. (l11);
  \draw [->] (l6) ..controls (210.27bp,100.36bp) and (199.55bp,88.994bp)  .. (l2);
  \draw [->] (l13) ..controls (222.88bp,153.43bp) and (188.97bp,139.76bp)  .. (l14);
  \draw [->] (l11) ..controls (54.035bp,100.36bp) and (64.967bp,88.994bp)  .. (l8);
  \draw [->] (l5) ..controls (263.2bp,99.426bp) and (229.55bp,85.756bp)  .. (l2);
  \draw [->] (l4) ..controls (254bp,47.046bp) and (254bp,37.029bp)  .. (l1);
  \draw [->] (l8) ..controls (142.4bp,45.171bp) and (187.56bp,30.94bp)  .. (l1);
  \draw [->] (l7) ..controls (311.68bp,45.709bp) and (291.13bp,32.982bp)  .. (l1);
  \draw [->] (l13) ..controls (252.39bp,154.52bp) and (244.09bp,143.45bp)  .. (l6);
  \draw [->] (l15) ..controls (443.04bp,100.59bp) and (435.27bp,89.679bp)  .. (l10);
\end{scope}
\end{tikzpicture}

%% file: C4.tex

\begin{tikzpicture}[>=latex,line join=bevel,]
\begin{scope}
  \pgfsetstrokecolor{black}
  \definecolor{strokecol}{rgb}{1.0,1.0,1.0};
  \pgfsetstrokecolor{strokecol}
  \definecolor{fillcol}{rgb}{1.0,1.0,1.0};
  \pgfsetfillcolor{fillcol}
  \filldraw (0bp,0bp) -- (0bp,234bp) -- (720bp,234bp) -- (720bp,0bp) -- cycle;
\end{scope}
\begin{scope}
  \pgfsetstrokecolor{black}
  \definecolor{strokecol}{rgb}{1.0,1.0,1.0};
  \pgfsetstrokecolor{strokecol}
  \definecolor{fillcol}{rgb}{1.0,1.0,1.0};
  \pgfsetfillcolor{fillcol}
  \filldraw (0bp,0bp) -- (0bp,234bp) -- (720bp,234bp) -- (720bp,0bp) -- cycle;
\end{scope}
\begin{scope}
  \pgfsetstrokecolor{black}
  \definecolor{strokecol}{rgb}{1.0,1.0,1.0};
  \pgfsetstrokecolor{strokecol}
  \definecolor{fillcol}{rgb}{1.0,1.0,1.0};
  \pgfsetfillcolor{fillcol}
  \filldraw (0bp,0bp) -- (0bp,234bp) -- (720bp,234bp) -- (720bp,0bp) -- cycle;
\end{scope}
\begin{scope}
  \pgfsetstrokecolor{black}
  \definecolor{strokecol}{rgb}{1.0,1.0,1.0};
  \pgfsetstrokecolor{strokecol}
  \definecolor{fillcol}{rgb}{1.0,1.0,1.0};
  \pgfsetfillcolor{fillcol}
  \filldraw (0bp,0bp) -- (0bp,234bp) -- (720bp,234bp) -- (720bp,0bp) -- cycle;
\end{scope}
\begin{scope}
  \pgfsetstrokecolor{black}
  \definecolor{strokecol}{rgb}{1.0,1.0,1.0};
  \pgfsetstrokecolor{strokecol}
  \definecolor{fillcol}{rgb}{1.0,1.0,1.0};
  \pgfsetfillcolor{fillcol}
  \filldraw (0bp,0bp) -- (0bp,234bp) -- (720bp,234bp) -- (720bp,0bp) -- cycle;
\end{scope}
\begin{scope}
  \pgfsetstrokecolor{black}
  \definecolor{strokecol}{rgb}{1.0,1.0,1.0};
  \pgfsetstrokecolor{strokecol}
  \definecolor{fillcol}{rgb}{1.0,1.0,1.0};
  \pgfsetfillcolor{fillcol}
  \filldraw (0bp,0bp) -- (0bp,234bp) -- (720bp,234bp) -- (720bp,0bp) -- cycle;
\end{scope}
\begin{scope}
  \pgfsetstrokecolor{black}
  \definecolor{strokecol}{rgb}{1.0,1.0,1.0};
  \pgfsetstrokecolor{strokecol}
  \definecolor{fillcol}{rgb}{1.0,1.0,1.0};
  \pgfsetfillcolor{fillcol}
  \filldraw (0bp,0bp) -- (0bp,234bp) -- (720bp,234bp) -- (720bp,0bp) -- cycle;
\end{scope}
\begin{scope}
  \pgfsetstrokecolor{black}
  \definecolor{strokecol}{rgb}{1.0,1.0,1.0};
  \pgfsetstrokecolor{strokecol}
  \definecolor{fillcol}{rgb}{1.0,1.0,1.0};
  \pgfsetfillcolor{fillcol}
  \filldraw (0bp,0bp) -- (0bp,234bp) -- (720bp,234bp) -- (720bp,0bp) -- cycle;
\end{scope}
\begin{scope}
  \pgfsetstrokecolor{black}
  \definecolor{strokecol}{rgb}{1.0,1.0,1.0};
  \pgfsetstrokecolor{strokecol}
  \definecolor{fillcol}{rgb}{1.0,1.0,1.0};
  \pgfsetfillcolor{fillcol}
  \filldraw (0bp,0bp) -- (0bp,234bp) -- (720bp,234bp) -- (720bp,0bp) -- cycle;
\end{scope}
\begin{scope}
  \pgfsetstrokecolor{black}
  \definecolor{strokecol}{rgb}{1.0,1.0,1.0};
  \pgfsetstrokecolor{strokecol}
  \definecolor{fillcol}{rgb}{1.0,1.0,1.0};
  \pgfsetfillcolor{fillcol}
  \filldraw (0bp,0bp) -- (0bp,234bp) -- (720bp,234bp) -- (720bp,0bp) -- cycle;
\end{scope}
\begin{scope}
  \pgfsetstrokecolor{black}
  \definecolor{strokecol}{rgb}{1.0,1.0,1.0};
  \pgfsetstrokecolor{strokecol}
  \definecolor{fillcol}{rgb}{1.0,1.0,1.0};
  \pgfsetfillcolor{fillcol}
  \filldraw (0bp,0bp) -- (0bp,234bp) -- (720bp,234bp) -- (720bp,0bp) -- cycle;
\end{scope}
\begin{scope}
  \pgfsetstrokecolor{black}
  \definecolor{strokecol}{rgb}{1.0,1.0,1.0};
  \pgfsetstrokecolor{strokecol}
  \definecolor{fillcol}{rgb}{1.0,1.0,1.0};
  \pgfsetfillcolor{fillcol}
  \filldraw (0bp,0bp) -- (0bp,234bp) -- (720bp,234bp) -- (720bp,0bp) -- cycle;
\end{scope}
\begin{scope}
  \pgfsetstrokecolor{black}
  \definecolor{strokecol}{rgb}{1.0,1.0,1.0};
  \pgfsetstrokecolor{strokecol}
  \definecolor{fillcol}{rgb}{1.0,1.0,1.0};
  \pgfsetfillcolor{fillcol}
  \filldraw (0bp,0bp) -- (0bp,234bp) -- (720bp,234bp) -- (720bp,0bp) -- cycle;
\end{scope}
\begin{scope}
  \pgfsetstrokecolor{black}
  \definecolor{strokecol}{rgb}{1.0,1.0,1.0};
  \pgfsetstrokecolor{strokecol}
  \definecolor{fillcol}{rgb}{1.0,1.0,1.0};
  \pgfsetfillcolor{fillcol}
  \filldraw (0bp,0bp) -- (0bp,234bp) -- (720bp,234bp) -- (720bp,0bp) -- cycle;
\end{scope}
\begin{scope}
  \pgfsetstrokecolor{black}
  \definecolor{strokecol}{rgb}{1.0,1.0,1.0};
  \pgfsetstrokecolor{strokecol}
  \definecolor{fillcol}{rgb}{1.0,1.0,1.0};
  \pgfsetfillcolor{fillcol}
  \filldraw (0bp,0bp) -- (0bp,234bp) -- (720bp,234bp) -- (720bp,0bp) -- cycle;
\end{scope}
\begin{scope}
  \pgfsetstrokecolor{black}
  \definecolor{strokecol}{rgb}{1.0,1.0,1.0};
  \pgfsetstrokecolor{strokecol}
  \definecolor{fillcol}{rgb}{1.0,1.0,1.0};
  \pgfsetfillcolor{fillcol}
  \filldraw (0bp,0bp) -- (0bp,234bp) -- (720bp,234bp) -- (720bp,0bp) -- cycle;
\end{scope}
\begin{scope}
  \pgfsetstrokecolor{black}
  \definecolor{strokecol}{rgb}{1.0,1.0,1.0};
  \pgfsetstrokecolor{strokecol}
  \definecolor{fillcol}{rgb}{1.0,1.0,1.0};
  \pgfsetfillcolor{fillcol}
  \filldraw (0bp,0bp) -- (0bp,234bp) -- (720bp,234bp) -- (720bp,0bp) -- cycle;
\end{scope}
\begin{scope}
  \pgfsetstrokecolor{black}
  \definecolor{strokecol}{rgb}{1.0,1.0,1.0};
  \pgfsetstrokecolor{strokecol}
  \definecolor{fillcol}{rgb}{1.0,1.0,1.0};
  \pgfsetfillcolor{fillcol}
  \filldraw (0bp,0bp) -- (0bp,234bp) -- (720bp,234bp) -- (720bp,0bp) -- cycle;
\end{scope}
\begin{scope}
  \pgfsetstrokecolor{black}
  \definecolor{strokecol}{rgb}{1.0,1.0,1.0};
  \pgfsetstrokecolor{strokecol}
  \definecolor{fillcol}{rgb}{1.0,1.0,1.0};
  \pgfsetfillcolor{fillcol}
  \filldraw (0bp,0bp) -- (0bp,234bp) -- (720bp,234bp) -- (720bp,0bp) -- cycle;
\end{scope}
\begin{scope}
  \pgfsetstrokecolor{black}
  \definecolor{strokecol}{rgb}{1.0,1.0,1.0};
  \pgfsetstrokecolor{strokecol}
  \definecolor{fillcol}{rgb}{1.0,1.0,1.0};
  \pgfsetfillcolor{fillcol}
  \filldraw (0bp,0bp) -- (0bp,234bp) -- (720bp,234bp) -- (720bp,0bp) -- cycle;
\end{scope}
\begin{scope}
  \pgfsetstrokecolor{black}
  \definecolor{strokecol}{rgb}{1.0,1.0,1.0};
  \pgfsetstrokecolor{strokecol}
  \definecolor{fillcol}{rgb}{1.0,1.0,1.0};
  \pgfsetfillcolor{fillcol}
  \filldraw (0bp,0bp) -- (0bp,234bp) -- (720bp,234bp) -- (720bp,0bp) -- cycle;
\end{scope}
\begin{scope}
  \pgfsetstrokecolor{black}
  \definecolor{strokecol}{rgb}{1.0,1.0,1.0};
  \pgfsetstrokecolor{strokecol}
  \definecolor{fillcol}{rgb}{1.0,1.0,1.0};
  \pgfsetfillcolor{fillcol}
  \filldraw (0bp,0bp) -- (0bp,234bp) -- (720bp,234bp) -- (720bp,0bp) -- cycle;
\end{scope}
\begin{scope}[rounded corners=2pt,semithick]
  \node (l18) at (54bp,171bp) [draw,rectangle] {$A_{3}^{2}E_{8}^{\text{st}}\;\; (4^{2})$};
  \node (l19) at (261bp,225bp) [draw,rectangle] {$D_{6}E_{8}^{\text{st}}\;\; (2^{2})$};
  \node (l14) at (525bp,171bp) [draw,rectangle] {$A_{1}^{2}A_{3}E_{7}^{\text{st}}\;\; (6^{2})$};
  \node (l15) at (136bp,171bp) [draw,rectangle] {$A_{1}A_{3}^{2}E_{7}^{\text{st}}\;\; (4^{2})$};
  \node (l16) at (261bp,171bp) [draw,rectangle] {$A_{1}^{3}D_{4}E_{7}^{\text{st}}\;\; (2^{4})$};
  \node (l17) at (438bp,171bp) [draw,rectangle] {$A_{1}^{2}D_{4}E_{8}^{\text{st}}\;\; (2^{4})$};
  \node (l10) at (195bp,63bp) [draw,rectangle] {$A_{1}E_{6}^{\text{st}}\;\; (2^{2}8^{2})$};
  \node (l11) at (402bp,63bp) [draw,rectangle] {$A_{1}^{2}E_{6}^{\text{st}}\;\; (2^{4}4^{2})$};
  \node (l12) at (558bp,63bp) [draw,rectangle] {$A_{1}^{3}E_{7}^{\text{st}}\;\; (2^{4}4^{2})$};
  \node (l13) at (438bp,117bp) [draw,rectangle] {$A_{1}^{7}E_{7}^{\text{st}}\;\; (2^{6})$};
  \node (l21) at (280bp,117bp) [draw,rectangle] {$A_{3}E_{6}^{\text{st}}\;\; (2^{6})$};
  \node (l20) at (523bp,117bp) [draw,rectangle] {$A_{1}A_{3}E_{7}^{\text{st}}\;\; (2^{2}4^{2})$};
  \node (l22) at (692bp,117bp) [draw,rectangle] {$A_{1}^{6}E_{8}^{\text{st}}\;\; (2^{6})$};
  \node (l6) at (359bp,117bp) [draw,rectangle] {$A_{1}^{5}E_{7}^{\text{st}}\;\; (2^{2}4^{2})$};
  \node (l7) at (33bp,117bp) [draw,rectangle] {$A_{1}^{5}E_{7}^{\text{st}}\;\; (2^{2}4^{2})$};
  \node (l4) at (324bp,63bp) [draw,rectangle] {$E_{6}^{\text{st}}\;\; (2^{4}4^{2})$};
  \node (l5) at (613bp,117bp) [draw,rectangle] {$A_{1}^{6}E_{7}^{\text{st}}\;\; (2^{3}8^{1})$};
  \node (l2) at (117bp,63bp) [draw,rectangle] {$E_{6}^{\text{st}}\;\; (4^{4})$};
  \node (l3) at (480bp,9bp) [draw,rectangle] {$E_{6}^{\text{st}}\;\; (2^{2}4^{4})$};
  \node (l1) at (480bp,63bp) [draw,rectangle] {$E_{6}^{\text{st}}\;\; (2^{2}8^{2})$};
  \node (l8) at (117bp,117bp) [draw,rectangle] {$A_{1}^{4}E_{6}^{\text{st}}\;\; (2^{2}4^{2})$};
  \node (l9) at (201bp,117bp) [draw,rectangle] {$A_{1}^{6}E_{6}^{\text{st}}\;\; (2^{2}4^{2})$};
\end{scope}
\begin{scope}[semithick]
  \draw [->] (l16) ..controls (242.7bp,154.14bp) and (229.04bp,142.3bp)  .. (l9);
  \draw [->] (l19) ..controls (196.01bp,207.67bp) and (132.09bp,191.62bp)  .. (l18);
  \draw [->] (l19) ..controls (318.41bp,207.13bp) and (367.14bp,192.82bp)  .. (l17);
  \draw [->] (l8) ..controls (117bp,101.05bp) and (117bp,91.029bp)  .. (l2);
  \draw [->] (l6) ..controls (397.48bp,99.464bp) and (429.09bp,85.877bp)  .. (l1);
  \draw [->] (l16) ..controls (318.41bp,153.13bp) and (367.14bp,138.82bp)  .. (l13);
  \draw [->] (l15) ..controls (209.49bp,152.86bp) and (273.49bp,137.94bp)  .. (l6);
  \draw [->] (l9) ..controls (266.64bp,99.019bp) and (322.97bp,84.445bp)  .. (l11);
  \draw [->] (l13) ..controls (476.07bp,99.502bp) and (507.24bp,85.998bp)  .. (l12);
  \draw [->] (l21) ..controls (318.8bp,99.464bp) and (350.67bp,85.877bp)  .. (l11);
  \draw [->] (l19) ..controls (261bp,209.05bp) and (261bp,199.03bp)  .. (l16);
  \draw [->] (l17) ..controls (494.76bp,153.13bp) and (542.94bp,138.82bp)  .. (l5);
  \draw [->] (l9) ..controls (199.27bp,101.05bp) and (198.12bp,91.029bp)  .. (l10);
  \draw [->] (l17) ..controls (438bp,155.05bp) and (438bp,145.03bp)  .. (l13);
  \draw [->] (l15) ..controls (130.48bp,154.89bp) and (126.71bp,144.58bp)  .. (l8);
  \draw [->] (l13) ..controls (401.62bp,99.403bp) and (372.28bp,86.025bp)  .. (l4);
  \draw [->] (l9) ..controls (174.68bp,99.709bp) and (154.13bp,86.982bp)  .. (l2);
  \draw [->] (l17) ..controls (413.37bp,153.79bp) and (394.28bp,141.22bp)  .. (l6);
  \draw [->] (l19) ..controls (221.25bp,207.46bp) and (188.59bp,193.88bp)  .. (l15);
  \draw [->] (l17) ..controls (464.63bp,153.71bp) and (485.43bp,140.98bp)  .. (l20);
  \draw [->] (l13) ..controls (427.33bp,100.59bp) and (419.78bp,89.679bp)  .. (l11);
  \draw [->] (l7) ..controls (85.308bp,99.21bp) and (129.38bp,85.063bp)  .. (l10);
  \draw [->] (l16) ..controls (186.12bp,152.92bp) and (119.67bp,137.77bp)  .. (l7);
  \draw [->] (l16) ..controls (266.52bp,154.89bp) and (270.29bp,144.58bp)  .. (l21);
  \draw [->] (l7) ..controls (59.317bp,99.709bp) and (79.874bp,86.982bp)  .. (l2);
  \draw [->] (l22) ..controls (649.19bp,99.387bp) and (613.75bp,85.634bp)  .. (l12);
  \draw [->] (l20) ..controls (484.52bp,99.464bp) and (452.91bp,85.877bp)  .. (l11);
  \draw [->] (l8) ..controls (186.13bp,98.633bp) and (247.65bp,83.181bp)  .. (l4);
  \draw [->] (l20) ..controls (533.37bp,100.59bp) and (540.72bp,89.679bp)  .. (l12);
  \draw [->] (l11) ..controls (426.21bp,45.861bp) and (444.82bp,33.457bp)  .. (l3);
  \draw [->] (l19) ..controls (335.11bp,209.4bp) and (426.11bp,191.48bp)  .. (l14);
  \draw [->] (l16) ..controls (214.79bp,153.31bp) and (176.23bp,139.39bp)  .. (l8);
  \draw [->] (l16) ..controls (340.77bp,154.17bp) and (421.48bp,138.15bp)  .. (l20);
  \draw [->] (l18) ..controls (47.9bp,154.89bp) and (43.735bp,144.58bp)  .. (l7);
  \draw [->] (l17) ..controls (513.53bp,154.97bp) and (583.97bp,140.82bp)  .. (l22);
  \draw [->] (l15) ..controls (103.35bp,153.52bp) and (77.34bp,140.39bp)  .. (l7);
  \draw [->] (l12) ..controls (533.79bp,45.861bp) and (515.18bp,33.457bp)  .. (l3);
  \draw [->] (l5) ..controls (570.51bp,99.387bp) and (535.33bp,85.634bp)  .. (l1);
  \draw [->] (l6) ..controls (348.63bp,100.59bp) and (341.28bp,89.679bp)  .. (l4);
\end{scope}
\end{tikzpicture}

%% file: C9.tex

\begin{tikzpicture}[>=latex,line join=bevel,]
\begin{scope}
  \pgfsetstrokecolor{black}
  \definecolor{strokecol}{rgb}{1.0,1.0,1.0};
  \pgfsetstrokecolor{strokecol}
  \definecolor{fillcol}{rgb}{1.0,1.0,1.0};
  \pgfsetfillcolor{fillcol}
  \filldraw (0bp,0bp) -- (0bp,72bp) -- (54bp,72bp) -- (54bp,0bp) -- cycle;
\end{scope}
\begin{scope}
  \pgfsetstrokecolor{black}
  \definecolor{strokecol}{rgb}{1.0,1.0,1.0};
  \pgfsetstrokecolor{strokecol}
  \definecolor{fillcol}{rgb}{1.0,1.0,1.0};
  \pgfsetfillcolor{fillcol}
  \filldraw (0bp,0bp) -- (0bp,72bp) -- (54bp,72bp) -- (54bp,0bp) -- cycle;
\end{scope}
\begin{scope}[rounded corners=2pt,semithick]
  \node (l2) at (27bp,63bp) [draw,rectangle] {$A_{6}E_{8}^{\text{st}}\;\; (7^{1})$};
  \node (l1) at (27bp,9bp) [draw,rectangle] {$E_{6}^{\text{st}}\;\; (7^{3})$};
\end{scope}
\begin{scope}[semithick]
  \draw [->] (l2) ..controls (27bp,47.046bp) and (27bp,37.029bp)  .. (l1);
\end{scope}
\end{tikzpicture}

%% file: C7.tex

\begin{tikzpicture}[>=latex,line join=bevel,]
\begin{scope}
  \pgfsetstrokecolor{black}
  \definecolor{strokecol}{rgb}{1.0,1.0,1.0};
  \pgfsetstrokecolor{strokecol}
  \definecolor{fillcol}{rgb}{1.0,1.0,1.0};
  \pgfsetfillcolor{fillcol}
  \filldraw (0bp,0bp) -- (0bp,180bp) -- (270bp,180bp) -- (270bp,0bp) -- cycle;
\end{scope}
\begin{scope}
  \pgfsetstrokecolor{black}
  \definecolor{strokecol}{rgb}{1.0,1.0,1.0};
  \pgfsetstrokecolor{strokecol}
  \definecolor{fillcol}{rgb}{1.0,1.0,1.0};
  \pgfsetfillcolor{fillcol}
  \filldraw (0bp,0bp) -- (0bp,180bp) -- (270bp,180bp) -- (270bp,0bp) -- cycle;
\end{scope}
\begin{scope}
  \pgfsetstrokecolor{black}
  \definecolor{strokecol}{rgb}{1.0,1.0,1.0};
  \pgfsetstrokecolor{strokecol}
  \definecolor{fillcol}{rgb}{1.0,1.0,1.0};
  \pgfsetfillcolor{fillcol}
  \filldraw (0bp,0bp) -- (0bp,180bp) -- (270bp,180bp) -- (270bp,0bp) -- cycle;
\end{scope}
\begin{scope}
  \pgfsetstrokecolor{black}
  \definecolor{strokecol}{rgb}{1.0,1.0,1.0};
  \pgfsetstrokecolor{strokecol}
  \definecolor{fillcol}{rgb}{1.0,1.0,1.0};
  \pgfsetfillcolor{fillcol}
  \filldraw (0bp,0bp) -- (0bp,180bp) -- (270bp,180bp) -- (270bp,0bp) -- cycle;
\end{scope}
\begin{scope}
  \pgfsetstrokecolor{black}
  \definecolor{strokecol}{rgb}{1.0,1.0,1.0};
  \pgfsetstrokecolor{strokecol}
  \definecolor{fillcol}{rgb}{1.0,1.0,1.0};
  \pgfsetfillcolor{fillcol}
  \filldraw (0bp,0bp) -- (0bp,180bp) -- (270bp,180bp) -- (270bp,0bp) -- cycle;
\end{scope}
\begin{scope}
  \pgfsetstrokecolor{black}
  \definecolor{strokecol}{rgb}{1.0,1.0,1.0};
  \pgfsetstrokecolor{strokecol}
  \definecolor{fillcol}{rgb}{1.0,1.0,1.0};
  \pgfsetfillcolor{fillcol}
  \filldraw (0bp,0bp) -- (0bp,180bp) -- (270bp,180bp) -- (270bp,0bp) -- cycle;
\end{scope}
\begin{scope}
  \pgfsetstrokecolor{black}
  \definecolor{strokecol}{rgb}{1.0,1.0,1.0};
  \pgfsetstrokecolor{strokecol}
  \definecolor{fillcol}{rgb}{1.0,1.0,1.0};
  \pgfsetfillcolor{fillcol}
  \filldraw (0bp,0bp) -- (0bp,180bp) -- (270bp,180bp) -- (270bp,0bp) -- cycle;
\end{scope}
\begin{scope}[rounded corners=2pt,semithick]
  \node (l6) at (135bp,171bp) [draw,rectangle] {$A_{1}D_{5}E_{8}^{\text{st}}\;\; (2^{1}4^{1})$};
  \node (l7) at (231bp,117bp) [draw,rectangle] {$A_{1}^{3}A_{3}E_{8}^{\text{st}}\;\; (2^{3}4^{1})$};
  \node (l4) at (39bp,117bp) [draw,rectangle] {$A_{1}^{3}A_{3}E_{7}^{\text{st}}\;\; (4^{1}8^{1})$};
  \node (l5) at (181bp,63bp) [draw,rectangle] {$A_{1}^{4}E_{7}^{\text{st}}\;\; (2^{1}4^{3})$};
  \node (l2) at (181bp,9bp) [draw,rectangle] {$E_{6}^{\text{st}}\;\; (2^{1}4^{1}8^{2})$};
  \node (l3) at (135bp,117bp) [draw,rectangle] {$A_{1}^{4}A_{3}E_{7}^{\text{st}}\;\; (2^{3}4^{1})$};
  \node (l1) at (89bp,63bp) [draw,rectangle] {$A_{1}^{2}E_{6}^{\text{st}}\;\; (2^{2}4^{1}8^{1})$};
\end{scope}
\begin{scope}[semithick]
  \draw [->] (l4) ..controls (54.035bp,100.36bp) and (64.967bp,88.994bp)  .. (l1);
  \draw [->] (l6) ..controls (165.22bp,153.63bp) and (189.01bp,140.74bp)  .. (l7);
  \draw [->] (l7) ..controls (215.96bp,100.36bp) and (205.03bp,88.994bp)  .. (l5);
  \draw [->] (l3) ..controls (121.23bp,100.44bp) and (111.31bp,89.223bp)  .. (l1);
  \draw [->] (l6) ..controls (135bp,155.05bp) and (135bp,145.03bp)  .. (l3);
  \draw [->] (l5) ..controls (181bp,47.046bp) and (181bp,37.029bp)  .. (l2);
  \draw [->] (l6) ..controls (104.78bp,153.63bp) and (80.989bp,140.74bp)  .. (l4);
  \draw [->] (l3) ..controls (148.77bp,100.44bp) and (158.69bp,89.223bp)  .. (l5);
\end{scope}
\end{tikzpicture}

%% file: C3.tex

\begin{tikzpicture}[>=latex,line join=bevel,]
\begin{scope}
  \pgfsetstrokecolor{black}
  \definecolor{strokecol}{rgb}{1.0,1.0,1.0};
  \pgfsetstrokecolor{strokecol}
  \definecolor{fillcol}{rgb}{1.0,1.0,1.0};
  \pgfsetfillcolor{fillcol}
  \filldraw (0bp,0bp) -- (0bp,126bp) -- (258bp,126bp) -- (258bp,0bp) -- cycle;
\end{scope}
\begin{scope}
  \pgfsetstrokecolor{black}
  \definecolor{strokecol}{rgb}{1.0,1.0,1.0};
  \pgfsetstrokecolor{strokecol}
  \definecolor{fillcol}{rgb}{1.0,1.0,1.0};
  \pgfsetfillcolor{fillcol}
  \filldraw (0bp,0bp) -- (0bp,126bp) -- (258bp,126bp) -- (258bp,0bp) -- cycle;
\end{scope}
\begin{scope}
  \pgfsetstrokecolor{black}
  \definecolor{strokecol}{rgb}{1.0,1.0,1.0};
  \pgfsetstrokecolor{strokecol}
  \definecolor{fillcol}{rgb}{1.0,1.0,1.0};
  \pgfsetfillcolor{fillcol}
  \filldraw (0bp,0bp) -- (0bp,126bp) -- (258bp,126bp) -- (258bp,0bp) -- cycle;
\end{scope}
\begin{scope}
  \pgfsetstrokecolor{black}
  \definecolor{strokecol}{rgb}{1.0,1.0,1.0};
  \pgfsetstrokecolor{strokecol}
  \definecolor{fillcol}{rgb}{1.0,1.0,1.0};
  \pgfsetfillcolor{fillcol}
  \filldraw (0bp,0bp) -- (0bp,126bp) -- (258bp,126bp) -- (258bp,0bp) -- cycle;
\end{scope}
\begin{scope}
  \pgfsetstrokecolor{black}
  \definecolor{strokecol}{rgb}{1.0,1.0,1.0};
  \pgfsetstrokecolor{strokecol}
  \definecolor{fillcol}{rgb}{1.0,1.0,1.0};
  \pgfsetfillcolor{fillcol}
  \filldraw (0bp,0bp) -- (0bp,126bp) -- (258bp,126bp) -- (258bp,0bp) -- cycle;
\end{scope}
\begin{scope}
  \pgfsetstrokecolor{black}
  \definecolor{strokecol}{rgb}{1.0,1.0,1.0};
  \pgfsetstrokecolor{strokecol}
  \definecolor{fillcol}{rgb}{1.0,1.0,1.0};
  \pgfsetfillcolor{fillcol}
  \filldraw (0bp,0bp) -- (0bp,126bp) -- (258bp,126bp) -- (258bp,0bp) -- cycle;
\end{scope}
\begin{scope}[rounded corners=2pt,semithick]
  \node (l6) at (123bp,117bp) [draw,rectangle] {$A_{2}D_{4}E_{8}^{\text{st}}\;\; (2^{1}6^{1})$};
  \node (l4) at (33bp,63bp) [draw,rectangle] {$A_{1}^{3}E_{7}^{\text{st}}\;\; (3^{1}6^{2})$};
  \node (l5) at (219bp,63bp) [draw,rectangle] {$A_{1}^{4}A_{2}E_{8}^{\text{st}}\;\; (2^{3}6^{1})$};
  \node (l2) at (194bp,9bp) [draw,rectangle] {$A_{1}A_{2}E_{7}^{\text{st}}\;\; (2^{2}4^{1}12^{1})$};
  \node (l3) at (123bp,63bp) [draw,rectangle] {$A_{1}^{5}A_{2}E_{7}^{\text{st}}\;\; (2^{3}6^{1})$};
  \node (l1) at (78bp,9bp) [draw,rectangle] {$E_{6}^{\text{st}}\;\; (2^{1}6^{3})$};
\end{scope}
\begin{scope}[semithick]
  \draw [->] (l4) ..controls (46.466bp,46.439bp) and (56.172bp,35.223bp)  .. (l1);
  \draw [->] (l3) ..controls (144.87bp,45.984bp) and (161.46bp,33.836bp)  .. (l2);
  \draw [->] (l3) ..controls (109.53bp,46.439bp) and (99.828bp,35.223bp)  .. (l1);
  \draw [->] (l6) ..controls (123bp,101.05bp) and (123bp,91.029bp)  .. (l3);
  \draw [->] (l6) ..controls (153.22bp,99.632bp) and (177.01bp,86.744bp)  .. (l5);
  \draw [->] (l5) ..controls (211.7bp,46.819bp) and (206.67bp,36.358bp)  .. (l2);
  \draw [->] (l6) ..controls (94.804bp,99.709bp) and (72.777bp,86.982bp)  .. (l4);
\end{scope}
\end{tikzpicture}

%% file: C2.tex

\begin{tikzpicture}[>=latex,line join=bevel,]
\begin{scope}
  \pgfsetstrokecolor{black}
  \definecolor{strokecol}{rgb}{1.0,1.0,1.0};
  \pgfsetstrokecolor{strokecol}
  \definecolor{fillcol}{rgb}{1.0,1.0,1.0};
  \pgfsetfillcolor{fillcol}
  \filldraw (0bp,0bp) -- (0bp,72bp) -- (68bp,72bp) -- (68bp,0bp) -- cycle;
\end{scope}
\begin{scope}
  \pgfsetstrokecolor{black}
  \definecolor{strokecol}{rgb}{1.0,1.0,1.0};
  \pgfsetstrokecolor{strokecol}
  \definecolor{fillcol}{rgb}{1.0,1.0,1.0};
  \pgfsetfillcolor{fillcol}
  \filldraw (0bp,0bp) -- (0bp,72bp) -- (68bp,72bp) -- (68bp,0bp) -- cycle;
\end{scope}
\begin{scope}[rounded corners=2pt,semithick]
  \node (l2) at (34bp,63bp) [draw,rectangle] {$A_{1}^{2}A_{2}^{2}E_{8}^{\text{st}}\;\; (6^{2})$};
  \node (l1) at (34bp,9bp) [draw,rectangle] {$A_{1}^{2}E_{7}^{\text{st}}\;\; (3^{2}6^{2})$};
\end{scope}
\begin{scope}[semithick]
  \draw [->] (l2) ..controls (34bp,47.046bp) and (34bp,37.029bp)  .. (l1);
\end{scope}
\end{tikzpicture}